\documentstyle[astrobib,epsf,times]{mn}
\begin{document}

\title[The Sunyaev--Zel'dovich angular power spectrum]
 {The Sunyaev--Zel'dovich angular power spectrum as a probe of cosmological
parameters}

\author[E. Komatsu \& U. Seljak]
 {E.~Komatsu$^1$\thanks{E-mail: komatsu@astro.princeton.edu} 
 and U.~Seljak$^2$\thanks{E-mail: useljak@princeton.edu} \\
 $^1$Department of Astrophysical Sciences, Princeton University,
 Princeton, NJ 08544, USA\\
 $^2$Department of Physics, Princeton University, Princeton, NJ 08544,
 USA}


\pagerange{\pageref{firstpage}--\pageref{lastpage}}
\pubyear{2002}

\maketitle

\label{firstpage}

\begin{abstract}
  The angular power spectrum of the Sunyaev--Zel'dovich (SZ) effect 
  is a powerful probe of cosmology. 
  It is easier to detect than individual clusters in the field, 
  is insensitive to observational selection
  effects and does not require a calibration between cluster mass and flux,
  reducing the systematic errors which dominate the
  cluster-counting constraints. It receives a dominant contribution from 
  virialized cluster region between $20-40\%$ of the virial radius
  and is thus relatively insensitive to the poorly known gas physics in the 
  cluster centre, such as cooling or (pre)heating.
  In this paper we derive a refined analytic prediction for the SZ angular 
  power spectrum using the universal gas-density and temperature profile
  and the dark-matter halo mass function.
  The predicted power spectrum has no free parameters and fits all 
  of the published hydrodynamic simulation results to better than
  a factor of two for $2000<l<10000$.   
  We find that the angular power spectrum $C_l$ scales as
  $C_l \propto \sigma_8^7(\Omega_{\rm b}h)^2$ and is almost independent of 
  all of the other cosmological parameters.
  This differs from the local cluster abundance studies, which give
  a relation between $\sigma_8$ and $\Omega_{\rm m}$.
  We also compute the covariance matrix of $C_l$ using the 
  halo model and find a good agreement relative to the simulations. 
  We argue that the best constraint from the SZ power spectrum 
  comes from $l \sim 3000$, where the sampling variance is sufficiently small 
  and the spectrum is dominated 
  by massive clusters above $10^{14}~h^{-1}~M_{\sun}$ for which cooling, 
  heating and details of star formation are not very important. 
  We estimate how well we can determine $\sigma_8(\Omega_{\rm b}h)^{2/7}$ with 
  sampling-variance-limited observations and find that for a 
  several-square-degree survey with arcminute resolution 
  one should be able to determine $\sigma_8$ to within a few percent, 
  with the remaining uncertainty dominated by theoretical modeling.
  If the recent excess of the CMB power on small scales reported by 
  CBI and BIMA experiments is due to the SZ effect, then we find 
  $\sigma_8(\Omega_{\rm b}h/0.035)^{0.29}=1.04 \pm 0.12$ at 95\% 
  confidence level (statistical)
  and with a residual 10\% systematic (theoretical) uncertainty.
\end{abstract}

\section{Introduction}\label{sec:intro}


Abundance of dark-matter halos and its redshift evolution have 
been recognized as an important cosmological test. 
For example, the local abundance of X-ray clusters can be used to constrain 
a combination of the matter-fluctuation amplitude and the matter density 
of the universe 
(e.g., \citeNP{1993MNRAS.262.1023W,1996MNRAS.281..323V,1996MNRAS.282..263E,1997ApJ...490..557K}). 
Higher-redshift clusters can be used to break this degeneracy
(e.g., \citeNP{1998ApJ...504....1B}), as well as constrain other cosmological 
parameters \cite{2001ApJ...553..545H}.

There are two major uncertainties in such analysis.
One is the sample itself, which has to be carefully constructed to avoid 
any possible selection biases that can affect the final constraint. 
The other is the relation between observable quantities 
(e.g., X-ray temperature or flux) and the halo mass, which is 
related to the theoretical models. 
The latter uncertainty is particularly important as a small error in this
relation can result in a significant effect on the final results. 
For example, the X-ray flux scales as gas-density 
squared and is dominated by the central regions, where complicated physical 
processes are taking place. 
This means that the relation between flux or flux-weighted
temperature and the halo mass is not easily obtained from {\it ab-initio} 
numerical simulations, and one must resort to more empirical calibrations
instead.

The Sunyaev--Zel'dovich (SZ) effect has some advantages over X-ray in this 
respect. 
The SZ effect is caused by the cosmic microwave background (CMB) photons 
scattering off hot electrons in the intracluster medium, and the 
SZ flux is proportional to the projected electron-gas pressure, which is 
proportional to a product 
of electron number density and temperature along the line of sight. 
Relative to X-ray emission 
this has a larger contribution from the outer parts of the cluster, where
the missing physics such as gas preheating or cooling in numerical simulations 
is less significant.

However, the SZ effect is very difficult to measure and currently 
only detections of already-known clusters exist \cite{1999PhR...310...97B}. 
It is likely that the first SZ detection in a random direction
will not be that of a specific cluster, but of a field correlation function,
or the angular power spectrum $C_l$, with several low signal-to-noise 
clusters contributing to the signal. 
Even if individual clusters are too faint to be detected with high statistical 
significance, the combined effect in $C_l$ can be statistically significant.
There have been several observational efforts in measuring the small-scale
fluctuations, and upper-limits have been obtained by 
ATCA \cite{1998MNRAS.298.1189S}, 
BIMA \cite{2000ApJ...539...57H,2001ApJ...553L...1D}, 
SUZIE \cite{1997ApJ...484..523C,1997ApJ...484..517G}, and 
VLA \cite{1997ApJ...483...38P}. 
Recently, CBI and BIMA experiments have reported statistically significant 
detections of the small-scale fluctuations at the level of 
$15-20~{\mu{\rm K}}$ on arcminutes angular scales 
\cite{astro-ph/0205384,2002astro.ph..6012D}, and their results may 
already be providing the first detection of the SZ fluctuations 
\cite{astro-ph/0205386}.

There are other advantages of using power spectrum in addition to 
the higher statistical significance in a noisy map. 
One is that selection effects are less important than in a cluster survey, 
for which flux, surface brightness, and other selection criteria have to be 
carefully examined to avoid any biases.
The other is that we do not have to measure the mass of individual haloes. 
As discussed above 
this is a dominant source of uncertainty in the local-mass-function 
determination from X-ray clusters and lack of a good calibration 
can lead to a 30\% error on $\sigma_8$. Even if mass were determined from SZ
flux, for which the central regions are less important, one would still need 
to include effects such as substructure, projections, ellipticity etc. 
As we argue in this paper using
the SZ power spectrum instead of the number-count analysis is less 
sensitive to selection biases and can still give statistically significant 
results. 
It is true that the power spectrum does not use all of the information 
available, specially if one obtains the information on cluster redshifts. 
However, the SZ power spectrum provides quite powerful 
constraints on cosmological parameters already by itself 
and the small increase in statistical errors is more than compensated
by the decrease in the systematic uncertainties.
Thus initial experimental efforts should focus on the power spectrum as a more 
reliable method to determine the cosmological parameters.

There are of course other effects that contribute to fluctuations 
in this range of wavelengths. 
The SZ effect is the dominant contribution for
$l>2000-4000$, depending on its amplitude.
The primary CMB dominates on the larger scales.
Although the primary CMB confuses the SZ effect, we can 
reduce or eliminate it by observing at 217~GHz at which the SZ effect 
vanishes. 
Many of the current generation CMB experiments (such as CBI) operate at low
frequencies, where the frequency dependence of CMB and the SZ effect is 
similar, and the two cannot be distinguished very well on this basis.
Galactic emission (synchrotron, dust, or free-free) does not have much 
power on the small angular scales, and has a different frequency dependence 
from CMB. They are not likely to be a major 
source of contamination. 
We ignore the kinetic SZ effect, as it is at least an order of magnitude
smaller in power than the thermal SZ effect.

Point sources have a lot of small-scale power and are potentially more
problematic contaminant.
While we can easily identify bright (more than several-$\sigma$) 
point sources, we cannot identify the fainter ones, which will contaminate 
the SZ effect.
This is particularly problematic for narrow frequency 
coverage experiments such as CBI, as we cannot use different 
frequency dependence of the point sources to separate the two components. 
We can still try to use opposite signatures between SZ and point sources
as a way to distinguish them; 
the SZ effect is negative in flux at low frequencies, while 
point sources are positive.
This results in a skewed one-point-distribution function. 
We can also use the difference in the shape of the correlation function, 
since more extended correlation in the SZ effect differs from the point 
sources. 
The SZ power spectrum scales roughly as $l^{-2}$ while that of the point 
sources is constant, so the two can be distinguished. 
In practice noise and beam smoothing may make this analysis more complicated.

Within a few years, several experiments will likely measure 
the SZ effect with an arcminute angular resolution and the effective area 
of several square degrees.
This has motivated us to revisit the predictions for the SZ angular power 
spectrum.
Many analytic calculations exist in the literature
\cite{1988MNRAS.233..637C,1993ApJ...405....1M,1999ApJ...515..465A,1999ApJ...526L...1K,2000PhRvD..62j3506C,2000ApJ...537..542M,2001ApJ...558..515H,2001ApJ...549...18Z}, as well as many hydrodynamic-simulation results
\cite{1993ApJ...416..399S,1995ApJ...442....1P,2000PhRvD..61l3001R,2001PhRvD..630f619S,2000MNRAS.317...37D,2001MNRAS.326..155D,2001ApJ...561L..15D,2001ApJ...549..681S,astro-ph/0012086,2002astro.ph..1375Z}.
The comparison between different simulations is difficult, as the parameters 
used differ, and the resulting power spectrum is extremely sensitive to the 
chosen parameters. 
Since massive clusters are at the tail of the dark-matter mass function,
the r.m.s. mass fluctuations within a $8~h^{-1}$~Mpc sphere,  
$\sigma_8$, is the most important parameter among the cosmological
parameters.
\citeN{1999ApJ...526L...1K} have shown that the matter density of 
the universe is less important than $\sigma_8$.

\citeN{2000PhRvD..61l3001R} and \citeN{astro-ph/0012086}
have attempted to compare analytic predictions based upon the so-called 
``halo approach'' (section~\ref{sec:cl}) with their simulated power 
spectrum, finding a good agreement. 
This agreement is, however, partially built into their model; in 
their approach, the parameters such as upper and lower mass cut-off are 
chosen to agree with the simulations. 
Another free parameter used in the analytic modeling that cannot be 
obtained from these models is the mass--temperature 
normalization.
Although a virial relation has often been used in this context,
the overall amplitude is treated as a free parameter, and
cannot be justified without a more detailed model. 
In this paper we develop a more refined analytic model and show that it
reproduces the simulation results without any free parameters, 
at least at the level of the simulations disagreeing among themselves.

The main ingredients into the halo model are the dark-matter-halo mass 
function, the gas-density and the gas-temperature profiles. 
For the latter, we use an analytic gas  and temperature profile model 
developed in our previous paper \cite{2001MNRAS.327.1353K}.
We have shown that it reproduces well 
the simulation results in the outer parts 
of the cluster which are relevant for the SZ effect.
This model fixes the normalization between mass and temperature by 
requiring that the gas and the dark-matter profiles agree with each other 
outside the gas core. 
Hydrodynamic simulations (e.g., \citeNP{1999ApJ...525..554F})
have shown 
that this assumption holds accurately.
As a result of this assumption and hydrostatic equilibrium
the gas temperature declines in the outer parts of the cluster, which 
is also seen in the simulations (although it remains controversial in 
X-ray observations due to the faint levels of X-ray emission in 
the outer parts of the cluster). 
This model therefore reproduces the main features of the hydrodynamic 
simulations and can be viewed as the gas analog of the often-used NFW 
profile \cite{1997ApJ...490..493N} for dark matter, at least outside the 
gas core where additional physics may play a role.

The structure of this paper is as follows. 
In \S~\ref{sec:cl} we outline the  ``halo approach'' to 
calculating the SZ angular power spectrum.
In \S~\ref{sec:profile} we derive the universal gas-density,
gas-temperature, and gas-pressure profiles following the prescription
described by \citeN{2001MNRAS.327.1353K}.
In \S~\ref{sec:simulations} we predict the spectrum and compare it to the
hydrodynamic simulations.
We then investigate the dependence of the spectrum
on various cosmological parameters.
We also study the mass and redshift distribution of the spectrum. 
In \S~\ref{sec:error} we compute the covariance matrix of $C_l$ 
by taking into account non-Gaussianity of the SZ fluctuations and
compare these to the hydrodynamic simulations.
In \S~\ref{sec:observations} we show how well we can
determine $\sigma_8$ with realistic SZ observations using
a likelihood analysis.
We put constraints on a $\Omega_{\rm m}-\sigma_8$ plane using the recent
CBI and BIMA data of the small-scale CMB fluctuations at $l\sim 2000-6000$ 
\cite{astro-ph/0205384,2002astro.ph..6012D}. 
In \S~\ref{sec:uncertainty} we quantify the theoretical uncertainty
in our predictions.
Finally, we summarize our results in \S~\ref{sec:conclusions}.
The fiducial cosmological model in this paper is
$\Omega_{\rm m}=0.37$, $\Omega_{\Lambda}=0.63$, $\Omega_{\rm b}=0.05$, 
$h=0.7$, $w=-1.0$, $n=1.0$, and $\sigma_8=1.0$.

\section{Halo Approach to the SZ Angular Power Spectrum}\label{sec:cl}

To compute the angular power spectrum of the SZ effect
we use the halo formalism 
\cite{1988MNRAS.233..637C,1993ApJ...405....1M,1999ApJ...515..465A,1999ApJ...526L...1K,2000PhRvD..62j3506C,2000ApJ...537..542M,2001ApJ...558..515H}.
For the angular scales of interest here,
$l>300$, the one-halo Poisson term dominates $C_l$ even after the 
subtraction of local massive haloes \cite{1999ApJ...526L...1K}.
We thus neglect the halo-halo correlation term throughout the paper.

The expression for $C_l$ is given by
\begin{equation}
 \label{eq:cl}
 C_l= g_\nu^2 \int_0^{z_{\rm max}} dz \frac{dV}{dz}
              \int_{M_{\rm min}}^{M_{\rm max}} dM \frac{dn(M,z)}{dM}
	      \left|\tilde{y}_l(M,z)\right|^2,
\label{halocl}
\end{equation}
where $g_\nu$ is the spectral function of the SZ effect
($-2$ in the Rayleigh--Jeans limit) \cite{1980ARA&A..18..537S},
$V(z)$ the comoving volume of the universe at $z$ per steradian, 
$dn(M,z)/dM$ the comoving dark-matter-halo mass function, and 
$\tilde{y}_l(M,z)$ the 2D Fourier transform of the projected 
Compton $y$-parameter.
The virial mass $M$ used here and its mass function are described 
in more detail in section~\ref{sec:dndM}.

For the upper integration boundary of redshift it suffices to take
$z_{\rm max}=10$ (figure~\ref{fig:dcdz}).
For the mass integration boundaries, $M_{\rm min}$ and $M_{\rm max}$,
we find that $M_{\rm min} = 5\times 10^{11}~h^{-1}~M_{\sun}$ and
$M_{\rm max} = 5\times 10^{15}~h^{-1}~M_{\sun}$ suffice to get 
the integral to converge on all angular scales, whereas
$M_{\rm min} = 5\times 10^{12}~h^{-1}~M_{\sun}$ is sufficient for $l<10^4$
(figure~\ref{fig:dcdM}).
Since the integrals converge the results are not subject to the specific
choice of integration boundaries.

The 2D Fourier transform of the projected Compton $y$-parameter,
$\tilde{y}_l= \tilde{y}_l(M,z)$, is given by
\begin{equation}
 \label{eq:yl}
  \tilde{y}_l= \frac{4\pi r_{\rm s}}{l_{\rm s}^2}
  \int_0^\infty dx x^2 y_{\rm 3D}(x)
  \frac{\sin(lx/l_{\rm s})}{lx/l_{\rm s}},
\end{equation}
where $y_{\rm 3D}(x)$ is the 3D radial profile of the Compton
$y$-parameter (equation~\ref{eq:p2y} in section~\ref{sec:profile}).
Note that $y_{\rm 3D}(x)$ has a dimension of $({\rm length})^{-1}$, 
while $\tilde{y}_l$ is dimensionless.
In equation~(\ref{eq:yl}) we have used $x$ as a scaled, non-dimensional 
radius, 
\begin{equation}
 \label{eq:x}
  x\equiv r/r_{\rm s},
\end{equation}
where $r_{\rm s}=r_{\rm s}(M,z)$ is  
a scale radius which characterizes the 3D radial profile and 
the corresponding 
angular wave number is 
\begin{equation}
 \label{eq:ls}
  l_{\rm s}\equiv d_{\rm A}/r_{\rm s}.
\end{equation}
Here $d_{\rm A}=d_{\rm A}(z)$ is the proper angular-diameter distance.
The projection of the 3D profile 
onto the sky in equation~(\ref{eq:yl})
was done within the Limber's approximation.
In section~\ref{sec:profile}, we derive $y_{\rm 3D}(x)$ using 
the universal gas and temperature profile.

\subsection{Dark-matter halo mass function}\label{sec:dndM}

For the dark-matter halo mass function $dn(M,z)/dM$ in equation (\ref{halocl})
we use  
the universal mass function \cite{2001MNRAS.321..372J}, which was 
derived from $N$-body simulations.
We should be careful about the definition of ``mass'' when using 
the mass function derived from $N$-body simulations 
(e.g., \citeNP{2001A&A...367...27W,2001astro.ph.11362S,2002astro.ph..3169H}).
The mass of haloes in the simulations is usually not the virial
mass $M$ as defined in our analytic formulation (equation \ref{eq:cl}).
Instead, the simulations usually use either the friend-of-friend (FOF) mass 
or the spherical overdensity (SO) mass to define the haloes.
We use here the SO mass, as the FOF mass is difficult to interpret in the
context of analytic formulation.

We define the SO mass, $M_\delta$, as the mass within the spherical
region whose density is $\delta$ times the critical density
of the universe at $z$ $\rho_{\rm c}(z)$
(note that sometimes one uses instead the mass within the spherical 
region whose density is $\tilde{\delta}$ times the mean mass density
of the universe at $z$, $\rho_{\rm m}(z)=\Omega_{\rm m}(z)\rho_{\rm c}(z)$;
the two are related through $\delta=\Omega_{\rm m}(z)\tilde{\delta}$).
Since $N$-body simulations give $dn(M_\delta,z)/dM_{\delta}$,
we need to convert it into $dn(M,z)/dM$.
We do this by calculating
\begin{equation}
 \label{eq:dndM}
  \frac{dn(M,z)}{dM}= \frac{dM_\delta}{dM}\frac{dn(M_\delta,z)}{dM_\delta}.
\end{equation}
For a given virial mass, $M$, we calculate $M_\delta$ with 
equation~(14) of \citeN{2001MNRAS.327.1353K}, and also evaluate 
$dM_\delta/dM$ numerically.

We use equation~(B3) of \citeN{2001MNRAS.321..372J},
which explicitly uses $\delta=180\Omega_{\rm m}(z)$
(mass density of haloes is 180 times the mean mass density of 
the universe at $z$),
\begin{eqnarray}
 \nonumber
  \frac{dn(M_{\delta},z)}{dM_{\delta}}
  &=& \Omega_{\rm m}(0)\frac{\rho_{\rm c}(0)}{M_{\delta}}
  \frac{d\ln\sigma^{-1}}{dM_{\delta}}\\
 \label{eq:dndM_jenkinsB3}
  & &\times 0.301\exp\left(-\left|\ln\sigma^{-1}+0.64\right|^{3.82}\right),
\end{eqnarray}
where $\sigma=\sigma(M_{\delta},z)$ is the linear, r.m.s. mass fluctuations
at a given redshift $z$ within the top-hat filter.
To compute $\sigma$ we use the BBKS transfer function 
\cite{1986ApJ...304...15B} with
the baryonic correction made by \citeN{1995ApJS..100..281S}.
The present-day critical density of the universe
is  $\rho_{\rm c}(0)=2.775\times 10^{11}~h^2~M_{\sun}~{\rm Mpc^{-3}}$.
Note that we have defined the mass function as the comoving number 
density.

While the functional formula for the mass function in
equation~(\ref{eq:dndM_jenkinsB3}) was tested for just one cosmological model
($\tau$CDM model) by \citeN{2001MNRAS.321..372J}, it is essentially 
a non-smoothed version of their universal mass function valid for a 
broad range of models. 
\citeN{2002astro.ph..3169H} have shown that the formula agrees 
with their simulated mass function as well which uses a different 
cosmological model.
In addition, \citeN{2002ApJ...573....7E} have shown that their mass function
for the SO mass haloes with $\delta=200$ agrees with the Jenkins et al.'s
formula, once taking into account the mass definition properly
(see also related 
discussion in \citeNP{2001astro.ph.11362S} and \citeNP{2002astro.ph..3169H}).
There is thus increasing evidence for the accuracy of 
Jenkins et al.'s mass function.

\section{gas-pressure profile of haloes}\label{sec:profile}

\subsection{Introduction}

The 3D Compton $y$-parameter profile, $y_{\rm 3D}(x)$, is given by a thermal 
gas-pressure profile, $P_{\rm gas}(x)$, through
\begin{eqnarray}
 \nonumber
  y_{\rm 3D}(x)
  &\equiv& \frac{\sigma_{\rm T}}{m_e c^2}P_e(x)
  = \frac{\sigma_{\rm T}}{m_e c^2}
  \left(\frac{2+2X}{3+5X}\right)P_{\rm gas}(x)\\
 \label{eq:p2y}
 &= &   
 1.04\times 10^{-4}~{\rm Mpc^{-1}}
 \left[\frac{P_{\rm gas}(x)}{50~{\rm eV~cm^{-3}}}\right],
\end{eqnarray}
where $P_e(x)$ is an electron-pressure profile, 
$X=0.76$ the primordial hydrogen abundance,
$\sigma_{\rm T}$ the Thomson cross section, 
$m_e$ the electron mass, and $c$ the speed of light.

We may write $P_{\rm gas}(x)$ with a gas-density profile, 
$\rho_{\rm gas}(x)$, and a gas-temperature profile, $T_{\rm gas}(x)$, as 
\begin{eqnarray}
 \nonumber
  P_{\rm gas}(x)&=& 
  \frac{3+5X}4
  \frac{\rho_{\rm gas}(x)}{m_p}
  k_{\rm B}T_{\rm gas}(x)\\
 \nonumber
  &=&
  55.0~{\rm eV~cm^{-3}}\\
 \label{eq:Pgas}
  & &\times
  \left[\frac{\rho_{\rm gas}(x)}{10^{14}~M_{\sun}~{\rm Mpc^{-3}}}\right]
  \left[\frac{k_{\rm B}T_{\rm gas}(x)}{8~{\rm keV}}\right],
\end{eqnarray}
where $m_p$ is the proton mass, and $k_{\rm B}$ the Boltzmann 
constant.

The aim of this section is to derive $P_{\rm gas}(x)$ 
following \citeN{2001MNRAS.327.1353K}.
Pressure profile was one of the uncertainties
in the previous work on the SZ power spectrum:
most of the previous work uses a spherical-isothermal $\beta$ model 
as a gas-density profile, with $\beta$ fixed at $2/3$ for 
simplicity \cite{1993ApJ...405....1M,1999ApJ...515..465A,1999ApJ...526L...1K,2000ApJ...537..542M,2001ApJ...558..515H};
\citeN{2000PhRvD..62j3506C} used an isothermal gas-density profile 
as predicted by \citeN{1998ApJ...497..555M}, which uses
hydrostatic equilibrium between gas pressure and dark-matter potential,
and is close to the $\beta$ profile.
The $\beta$ profile does not, however, reproduce the simulation results
in the outer parts of the haloes.
The outer slope of the $\beta$ model with $\beta=2/3$ scales as $r^{-2}$, 
being significantly shallower than the gas profile
in simulations which, as we have argued, is similar to the dark-matter 
profile and asymptotically scales as $r^{-3}$.
Since the SZ effect is sensitive to the outer parts of haloes,
this can cause a significant error in the predictions.

There is a related uncertainty caused by the shallow ($r^{-2}$) profile:
one has to cut out the extension of gas in haloes at
an arbitrary radius, otherwise one predicts too much power 
on large angular scales.
We find that the previous predictions are unstable against this 
cut-off radius.
This radius, which is a free parameter, has been assumed to 
be the virial radius; however, for isothermal gas assumed in the previous work 
this is {\it ad hoc}, as neither temperature nor density profile shows
the abrupt cut-off at the virial radius in hydrodynamic simulations 
\cite{1998ApJ...495...80B,1998ApJ...503..569E,1999ApJ...525..554F}.
Instead, both the temperature and the density decrease smoothly across the
virial radius, and it is only at $2-3$ times the virial radius that there 
seems to be an abrupt decrease of the temperature to the IGM value, 
the so-called shock radius \cite{1998ApJ...495...80B}.

Once the decline in profiles relative to the $\beta$ model is included, 
the contribution from the outer parts of haloes to the SZ power spectrum 
converges well within the shock radius, and is thus independent of the 
outer cut-off of the profile. 
This is because requiring that the gas-density
profile traces the dark-matter-density profile leads to
$\rho_{\rm gas}\propto r^{-3}$ in the outer parts of the haloes.
This by itself would give a logarithmically divergent gas-mass profile; 
however, since the temperature is also declining with radius,  
the resulting pressure profile is convergent.
Hydrostatic equilibrium and the gas density tracing the dark-matter density
make the gas temperature decrease with radius.

\subsection{Universal dark-matter-density profile}

The universal NFW dark-matter-density profile \cite{1997ApJ...490..493N} is
\begin{equation}
 \label{eq:NFWprofiles}
  \rho_{\rm dm}(x) = \frac{\rho_{\rm s}}{x(1+x)^2},
\end{equation}
where $x\equiv r/r_{\rm s}$, $r_{\rm s}$ a scale radius,
and $\rho_{\rm s}$ a scale density.
While the exact shape in the inner parts is still uncertain
\cite{1998ApJ...499L...5M,2000ApJ...529L..69J,2001ApJ...554..903K},
it does not affect the SZ effect, given the scales of interest here.

To specify the NFW profile, we need to 
specify a scale radius, $r_{\rm s}$, as a function of mass and redshift.
It is customary to specify the concentration parameter $c$ instead of 
$r_{\rm s}$,
\begin{equation}
 \label{eq:c}
  c(M,z)\equiv \frac{r_{\rm vir}(M,z)}{r_{\rm s}(M,z)}
  \approx \frac{10}{1+z}\left[\frac{M}{M_*(0)}\right]^{-0.2},
\end{equation}
where $r_{\rm vir}(M,z)$ is the virial radius of haloes, and the 
last expression follows from \citeN{2000MNRAS.318..203S}  
with redshift evolution found by \citeN{2001MNRAS.321..559B}.
The ``non-linear mass'' at $z=0$, $M_*(0)$, is a solution to
$\sigma(M)=\delta_c$, where $\sigma(M)$ is the present-day r.m.s. mass
fluctuations within the top-hat filter, $M$ the virial mass, 
and $\delta_c$ the threshold overdensity of spherical collapse at
$z=0$ \cite{1993MNRAS.262..627L,1997PThPh..97...49N}.
We calculate the virial radius using the spherical collapse model, 
\begin{equation}
 \label{eq:rvir}
  r_{\rm vir}(M,z)
  \equiv \left[
     \frac{M}
     {(4\pi/3)\Delta_{\rm c}(z)\rho_{\rm c}(z)}
   \right]^{1/3},
\end{equation}
where $\Delta_{\rm c}(z)$ is a spherical overdensity of
the virialized halo within $r_{\rm vir}$ at $z$, in units of 
the critical density of the universe at $z$
\cite{1993MNRAS.262..627L,1997PThPh..97...49N}.

The exact form of $c(M,z)$, equation~(\ref{eq:c}), is somewhat uncertain.
There are various fitting formulae in the literature
\cite{1997ApJ...490..493N,2001ApJ...554..114E,2001MNRAS.321..559B,2000MNRAS.318..203S,2000ApJ...535L...9C}.
In section~\ref{sec:simulations} we study the effect of different 
concentration parameters on our predictions, in addition to using our fiducial 
concentration parameter dependence in equation~(\ref{eq:c}).
We find that the difference in concentration parameters does not affect our 
predictions, largely because the SZ effect is not sensitive
to the central parts of haloes, where the profile shape makes most difference.

\subsection{Universal gas-pressure profile}

To derive a gas-pressure profile $P_{\rm gas}(x)$ 
we make three assumptions: (1) hydrostatic equilibrium between gas 
pressure and dark-matter potential due to the universal dark-matter
 density profile, 
(2) gas density tracing the universal dark-matter density in the 
outer parts of haloes, and 
(3) a constant polytropic equation of state for gas, 
$P_{\rm gas}\propto \rho_{\rm gas}^{\gamma}$.
The assumption (3) is obviously just an approximation and 
can be justified only to the extent that it suffices to explain 
the observations or simulations. 
There is evidence for this assumption to fail within inner 5\% of the virial 
radius, where X-ray observations indicate that temperature increases with 
radius \cite{2001MNRAS.328L..37A}. 
Outside the core region, nevertheless, the constant polytropic index 
seems to suffice, at least in comparison to the simulations.

From these three assumptions one can obtain universal gas-density, 
gas-temperature, and gas-pressure profiles as
\begin{equation}
 \label{eq:gas}
  \rho_{\rm gas}(x)= \rho_{\rm gas}(0)y_{\rm gas}(x),
\end{equation}
\begin{equation}
 \label{eq:temperature}
  T_{\rm gas}(x)= T_{\rm gas}(0)y^{\gamma-1}_{\rm gas}(x),
\end{equation}
and
\begin{equation}
 \label{eq:pressure}
  P_{\rm gas}(x)= P_{\rm gas}(0)y^\gamma_{\rm gas}(x).
\end{equation}
Solving the hydrostatic equilibrium equation with the {\it ansatz} above, 
we find that the non-dimensional gas-density profile, 
$y_{\rm gas}(x)$, has an analytic solution, and is given by
\cite{2001MNRAS.327.1353K}
\begin{equation}
 \label{eq:ygas}
  y_{\rm gas}(x)\equiv  
  \left\{
   1 - B\left[1-\frac{\ln(1+x)}x\right]
  \right\}^{1/\left(\gamma-1\right)},
\end{equation}
where the coefficient $B$ is
\begin{equation}
 \label{eq:B}
  B\equiv 
  3\eta^{-1}(0)\frac{\gamma-1}{\gamma}
    \left[\frac{\ln(1+c)}c-\frac1{1+c}\right]^{-1}.
\end{equation}
In appendix, we give exact formulae for the polytropic index,
$\gamma$, and the mass--temperature 
normalization factor at the centre, $\eta(0)$. 
Here we provide instead useful fitting formulae:
\begin{equation}
 \label{eq:gammafit}
  \gamma= 1.137 + 8.94\times 10^{-2}\ln(c/5) - 3.68\times 10^{-3}(c-5),
\end{equation}
and
\begin{equation}
 \label{eq:eta0fit}
  \eta(0)= 2.235 + 0.202(c-5) - 1.16\times 10^{-3}\left(c-5\right)^2.
\end{equation}
These fitting formulae are valid for $1< c < 25$.
Since $\gamma>1$ and the density decreases with radius, 
the temperature also decreases with radius.

From $\eta(0)$, we can determine the central temperature, $T_{\rm gas}(0)$, as
\begin{eqnarray}
 \nonumber
  T_{\rm gas}(0)
  &=& \eta(0)\frac{4}{3+5X}\frac{Gm_pM}{3r_{\rm vir}}\\
  \label{eq:Tgas0}
 &= &
 8.80~{\rm keV}~\eta(0)
 \left[
  \frac{M/(10^{15}~h^{-1}~M_{\sun})}{r_{\rm vir}/(1~h^{-1}~{\rm Mpc})}
 \right].
\end{eqnarray}
The central gas density, $\rho_{\rm gas}(0)$, is determined by
equation~(\ref{eq:rhog0}) below.
We then compute the central gas pressure, $P_{\rm gas}(0)$, by
substituting $\rho_{\rm gas}(0)$ and $T_{\rm gas}(0)$ into 
equation~(\ref{eq:Pgas}).
This completely specifies the gas-density, the gas-temperature,
and the gas-pressure profiles of haloes and their normalizations.

We determine the amount of gas in haloes by requiring that 
the gas density in the outer parts of haloes is 
$\Omega_{\rm b}/\Omega_{\rm m}$ times the dark-matter density.
Unlike the $\beta$ profile this is self-consistent,
as our gas-density profile agrees with the dark-matter profile outside 
the gas core. 
The virial radius thus has no special importance, and the 
matching can be performed over a broad range of radii without
affecting the results. We should reduce the gas fraction 
by the amount that has been converted to stars, which is of order 
$10-20\%$ for massive clusters of interest here \cite{1998ApJ...503..518F}. 
We choose not to apply this 
correction, so our predictions for the power spectrum are 
somewhat overestimated and the corresponding amplitude 
constraints underestimated. 
This however translates into only a few percent correction to the 
amplitude of $\sigma_8$.

We normalize the gas density at the virial radius,
\begin{eqnarray}
 \nonumber
  \rho_{\rm gas}(c)
  &=& \rho_{\rm gas}(0)y_{\rm gas}(c)
  = \frac{\Omega_{\rm b}}{\Omega_{\rm m}}\rho_{\rm dm}(c)\\
  \label{eq:norm}
  &=& \frac{\Omega_{\rm b}}{\Omega_{\rm m}}
  \frac{M}{4\pi r_{\rm vir}^3}\frac{c^2}{(1+c)^2}
  \left[\ln(1+c)-\frac{c}{1+c}\right]^{-1}.
\end{eqnarray}
We use this equation to obtain the central gas density, 
\begin{eqnarray}
 \nonumber
  \rho_{\rm gas}(0)
  &=& 7.96\times 10^{13}~M_{\sun}~{\rm Mpc^{-3}}\\
 \nonumber
  & &\times \left(\frac{\Omega_{\rm b}h^2}{\Omega_{\rm m}}\right)
  \frac{M/(10^{15}~h^{-1}~M_{\sun})}
  {\left[r_{\rm vir}/(1~h^{-1}~{\rm Mpc})\right]^3}\\
 \label{eq:rhog0}
 & &\times
 c^3\left[\frac{y_{\rm gas}^{-1}(c)}{c^2(1+c)^2}\right]
  \left[\ln(1+c)-\frac{c}{1+c}\right]^{-1}.
\end{eqnarray}

\section{Predictions for the SZ Angular Power Spectrum}
\label{sec:simulations}

In this section, we predict the SZ angular power spectrum, $C_l$, by
integrating equation~(\ref{eq:cl}), and compare the predictions with 
hydrodynamic simulations.
We then study the sensitivity of $C_l$ to various cosmological parameters.
We also study mass and redshift distribution of $C_l$,
finding that the power spectrum
primarily probes massive haloes in a moderate redshift 
universe.

\subsection{Comparison with hydrodynamic simulations}

\begin{figure}
  \begin{center}
    \leavevmode\epsfxsize=8.4cm \epsfbox{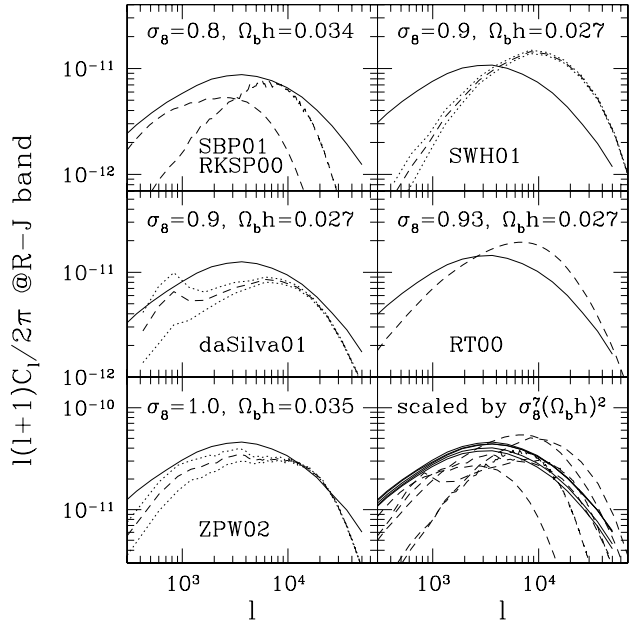}
  \end{center}
\caption[]
 {Comparison between the predicted SZ angular power spectra (solid)
 and the simulations (dashed; see table~\ref{tab:simulations} 
 for the meaning of the labels shown in each panel, and for the 
 cosmological parameters used).
 The dotted lines indicate r.m.s. errors of the simulated power spectra.
 The bottom-right panel scales all the power spectra by 
 $\sigma_8^7\left(\Omega_{\rm b}h\right)^2$. 
 The agreement between the scaled power spectra indicates that this 
 combination of cosmological parameters controls the amplitude of $C_l$.}
\label{fig:simulations}
\end{figure}

We begin by comparing our predictions for $C_l$ with 
six hydrodynamic simulations (four independent codes), listed in 
table~\ref{tab:simulations}.
Figure~\ref{fig:simulations} compares our analytic predictions
with the simulated power spectra.

The predictions generally agree with the simulations within a factor of two.
We find the best agreement for the nominally highest-resolution ($512^3$)
simulation run by \citeN{2002astro.ph..1375Z} using mesh-based 
MMH code (bottom-left panel of figure~\ref{fig:simulations}). 
The same code, but for different cosmological parameters and at  
lower ($256^3$) resolution, agrees well below $l<10000$, but lacks power 
above that \cite{2001PhRvD..630f619S} (top-left panel).
Note that the number of realizations for this simulation 
was small, which explains lack of power at low $l$ compared to 
even lower ($128^3$) resolution used by \citeN{2000PhRvD..61l3001R}
for the same code and the same cosmological parameters.

For the $256^3$ simulation we have created more 2-d maps,
covering the effective area of 1000 square degrees.
We find that with more realizations $C_l$ at low $l$ increases, 
becoming into better agreement with \shortciteN{2000PhRvD..61l3001R}.
This indicates that the sampling variance is very large on large scales, 
but is significantly reduced above $l \sim 2000-3000$. 
With more realizations, one typically increases power on large scales, 
which explains some of the differences among different simulations
(they differ in the effective simulated area significantly; see table~1).
Even so most of the simulations are too small to include sufficient number 
of very massive halos and for these 
multiple maps from a single simulation do not 
reduce the sampling variance. 
This can explain why the simulations are still below 
the analytic model at low $l$.
At least for MMH code, we find that the numerical resolution seems to be 
important at high $l$, and it causes some of the discrepancy 
between our model and the MMH simulations.

The comparison with one of the two SPH simulations,
\citeN{2001ApJ...561L..15D} using HYDRA code, shows a good agreement in shape,
while the GADGET code \cite{2001ApJ...549..681S} seems to 
predict more small-scale power above $l \sim 5000$ compared to
our predictions. 
We find that the shape of $C_l$ of \citeN{2001ApJ...549..681S} is 
similar to \citeN{astro-ph/0012086}, which uses a mesh-based hydrodynamic 
code, RAMSES, so the disagreement is unlikely to be due to simply 
SPH versus mesh-based codes. 
We note that the SPH simulation presented 
in \citeN{astro-ph/0205386} also agrees in shape with 
\citeN{2001ApJ...549..681S}.

A possible explanation for the discrepancy is that some simulations 
such as MMH have not yet converged for very high $l$, while at low 
$l$ sampling variance seems to be the most likely explanation.  
The discrepancy between our analytic model and GADGET and RAMSES could be 
caused by failure of our model at high $l$: clusters at higher redshift 
are less regular with more merging and substructure, which boosts $C_l$ 
at high $l$. One way to account for this in analytic model would be to  
include subhalos within halos, but to properly calibrate this one would 
need to use hydrodynamic simulations.
Here we argue that around $l \sim 3000$ the theoretical model and 
the simulations seem to agree rather well, certainly within a factor 
of two or better. Any remaining discrepancy at low $l$ could be due to 
insufficient sampling in simulations. 
Although a factor of two may sound like a large uncertainty in the $C_l$ 
prediction, it translates into a rather small uncertainty in $\sigma_8$, 
as discussed below.

\subsection{Dependence on cosmological parameters}

\begin{figure}
  \begin{center}
    \leavevmode\epsfxsize=8.4cm \epsfbox{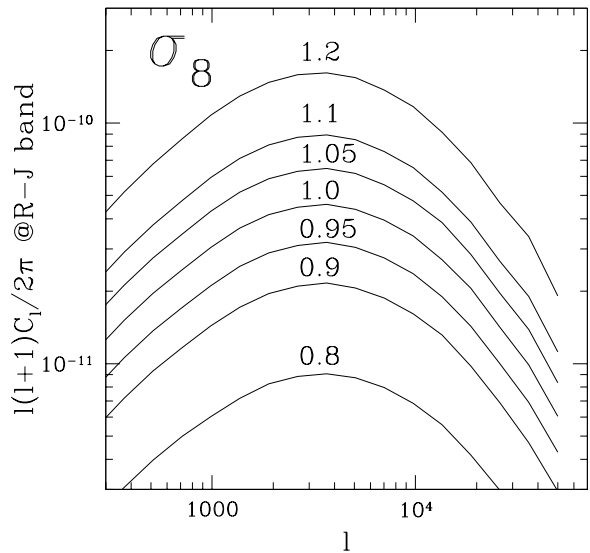}
  \end{center}
\caption[]
 {Dependence of the SZ angular power spectrum on $\sigma_8$.
 From top to bottom, the lines indicate 
 $\sigma_8=1.2$, 1.1, 1.05, 1.0, 0.95, 0.9, and 0.8, as shown in the figure.}
\label{fig:sigma8}
\end{figure}

In previous work it has been argued that the SZ power spectrum is very 
sensitive to $\sigma_8$ \cite{1999ApJ...526L...1K,2001PhRvD..630f619S,2001ApJ...549...18Z,2002astro.ph..1375Z}.
Figure~\ref{fig:sigma8} shows this dependence within the analytic model.
For example, a 40\% change in $\sigma_8$ causes one order of magnitude
change in $C_l$, while a 10\% change in $\sigma_8$ causes a factor of 
two change.
We find that $C_l \propto \sigma_8^7$ provides a good description of this 
scaling (see also the bottom-right panel of figure~\ref{fig:simulations}).
The strong dependence of $C_l$ on $\sigma_8$ indicates that 
even if the numerical simulations are uncertain at the level of a factor of 
two, this translates into less than 10\% systematic uncertainty in $\sigma_8$.

This property of the SZ power spectrum should be compared with the current 
uncertainty in $\sigma_8$ from the local cluster abundance studies, for which
the systematic uncertainty is larger than 10\% because of the uncertainty 
in the mass--temperature relation 
(e.g., \citeNP{2001MNRAS.325...77P,2001astro.ph.11362S,2002ApJ...569L..75V}).
Weak lensing (e.g., \citeNP{2002astro.ph..2503V})
and direct mass--temperature calibration 
may soon provide a more reliable method to determine 
$\sigma_8\Omega_{\rm m}^{0.5}$ combination.
On the other hand, we show below that $C_l$ is not sensitive 
to $\Omega_{\rm m}$ in the range of current interest.
This property can break the degeneracy between the two parameters.

\begin{figure}
  \begin{center}
    \leavevmode\epsfxsize=8.4cm \epsfbox{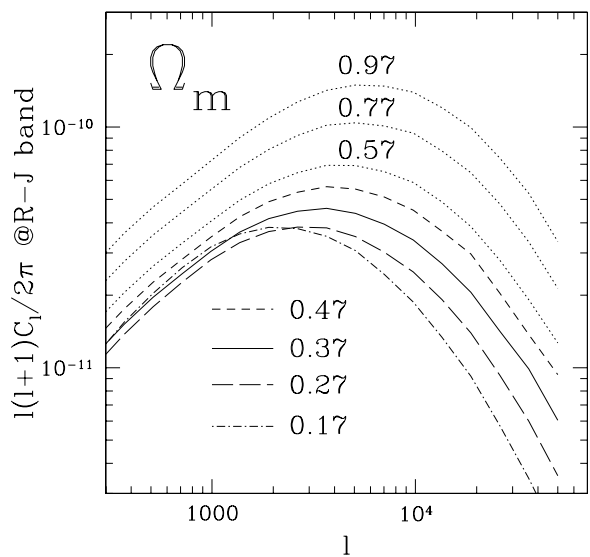}
  \end{center}
\caption[]
 {Dependence of the SZ angular power spectrum on $\Omega_{\rm m}$.
 From top to bottom, the lines indicate 
 $\Omega_{\rm m}=0.97$, 0.77, 0.57, 0.47, 0.37, 0.27, and 0.17, as shown in 
 the figure.
 While varying $\Omega_{\rm m}$, we have assumed a flat universe, 
 i.e., $\Omega_{\Lambda}=1-\Omega_{\rm m}$, and $w=-1.0$.}
\label{fig:omegam}
\end{figure}

Figure~\ref{fig:omegam} shows dependence of $C_l$ on $\Omega_{\rm m}$
assuming a flat universe, i.e., $\Omega_{\Lambda}=1-\Omega_{\rm m}$.
While $C_l\propto \Omega_{\rm m}$ for $\Omega_{\rm m}>0.4$,
we find that it is almost independent of $\Omega_{\rm m}$ for 
$0.15 < \Omega_{\rm m} < 0.4$.
This is because the comoving volume of the universe increases rapidly
with $\Omega_{\rm m}$ decreasing in this region, canceling out the effect
of $\Omega_{\rm m}$ on 
the dark-matter mass function.
As this range of $\Omega_{\rm m}$ is favored by current observations, $C_l$
is practically independent of $\Omega_{\rm m}$.

\begin{figure}
  \begin{center}
    \leavevmode\epsfxsize=8.4cm \epsfbox{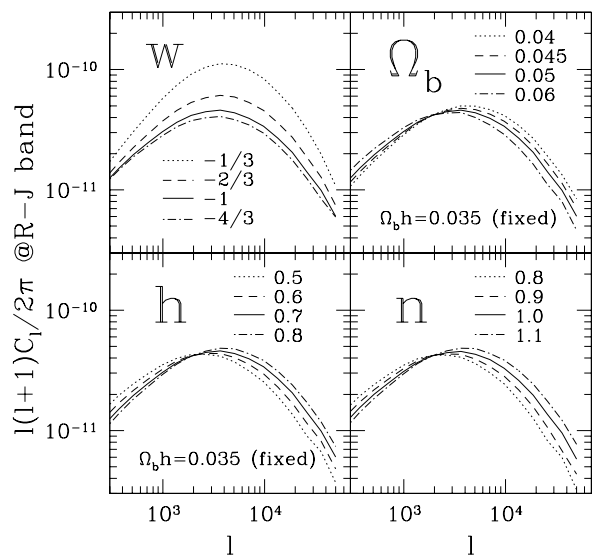}
  \end{center}
\caption[]
 {Dependence of the SZ angular power spectrum on the equation of state
 of the dark-energy component, $w$, the baryon density 
 $\Omega_{\rm b}$, the Hubble constant in units of 
 $100~{\rm km~s^{-1}~Mpc^{-1}}$, $h$, and the primordial 
 power-spectrum slope $n$.
 Values of each parameter are shown in each panel of the figure.
 While varying $\Omega_{\rm b}$ or $h$, we have $\Omega_{\rm b}h=0.035$ fixed.
 The panels for $\Omega_{\rm b}$ and $h$ are to show any residual
 dependence of $C_l$ on these two parameters.
 This figure is to be compared with figure~\ref{fig:sigma8}, showing
 how insensitive $C_l$ is to $w$, $\Omega_{\rm b}$, $h$,
 or $n$ compared to $\sigma_8$.}
\label{fig:cosmoparam}
\end{figure}

Figure~\ref{fig:cosmoparam} shows dependence of $C_l$ on equation of state $w$, 
$\Omega_{\rm b}$, $h$, and $n$.
While varying $\Omega_{\rm b}$ or $h$, we have fixed 
$\Omega_{\rm b}h$ at a fiducial value, 0.035, since we know that
$C_l$ scales as $(\Omega_{\rm b}h)^2$.  
The figure shows any residual dependence of $C_l$ on $\Omega_{\rm b}$ or $h$.
We find that $C_l$ depends upon $w$, $\Omega_{\rm b}$, $h$, or $n$
very weakly compared to $\sigma_8$ 
(compare figure~\ref{fig:cosmoparam} with figure~\ref{fig:sigma8}).
The only exception is if $w>-2/3$, which is however not favored by 
current observations \cite{2002PhRvD..65d1302B}.

We find that the dependence of overall amplitude of $C_l$
on cosmological parameters for $0.15<\Omega_{\rm m}<0.4$ is well 
approximated by
\begin{equation}
 \label{eq:fit}
 { l(l+1)C_l \over 2\pi}\simeq 330~\mu{\rm K}^2~\sigma_8^7 
  \left(\frac{\Omega_{\rm b}h}{0.035}\right)^2
  \qquad \left(0.15 < \Omega_{\rm m} < 0.4\right),
\end{equation}
at the peak around $l \sim 4000$.
As we have shown in figure~\ref{fig:omegam}, $C_l$ is rather insensitive 
to $\Omega_{\rm m}$ over this region.
A fit for $\Omega_{\rm m}>0.4$ is given by
\begin{equation}
  C_l\propto \left(\Omega_{\rm b}h\right)^2 \Omega_{\rm m} 
  \sigma_8^{6.5\Omega_{\rm m}^{-0.2}-0.9\sigma_8}
  \qquad \left(\Omega_{\rm m}> 0.4\right).
\end{equation}

\subsection{Redshift and mass distribution}

\begin{figure}
  \begin{center}
    \leavevmode\epsfxsize=8.4cm \epsfbox{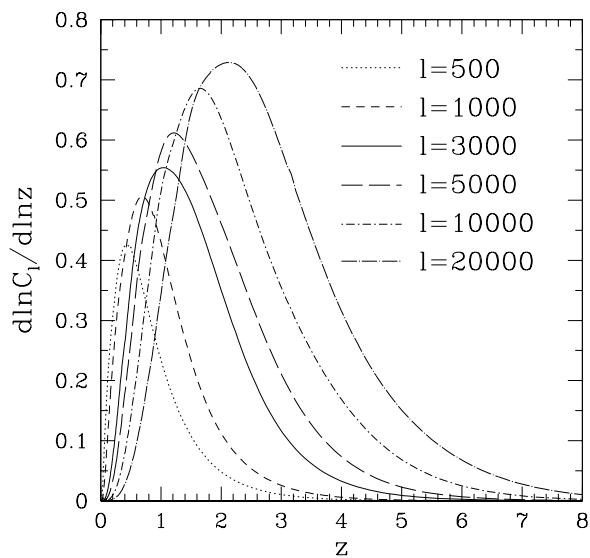}
  \end{center}
\caption[]
 {Redshift distribution of $C_l$.
 We plot $d\ln C_l/d\ln z$, equation~(\ref{eq:dlncldlnz}), for a given $l$.
 From left to right it is shown $l=500$, 1000, 3000, 5000, 10000,
 and 20000, as indicated in the figure.}
\label{fig:dcdz}
\end{figure}

Figure~\ref{fig:dcdz} shows the redshift distribution of $C_l$
for a given $l$, 
\begin{equation}
 \label{eq:dlncldlnz}
 \frac{d\ln C_l}{d\ln z}\equiv 
 \frac{z\frac{dV}{dz}
 \int dM \frac{dn(M,z)}{dM}
	      \left|\tilde{y}_l(M,z)\right|^2}
 {\int dz \frac{dV}{dz}
 \int dM \frac{dn(M,z)}{dM}
	      \left|\tilde{y}_l(M,z)\right|^2}.
\end{equation}
We find that haloes at $z \sim 1$ determine $C_l$ at $l \sim 3000$ 
(angular scales around $3'$).
Haloes at $z\sim 2$ dominate $C_l$ at $l=10000$.
Even haloes at $z\sim 3$ have a non-negligible contribution to $C_l$ 
for $l>10000$.

\begin{figure}
  \begin{center}
    \leavevmode\epsfxsize=8.4cm \epsfbox{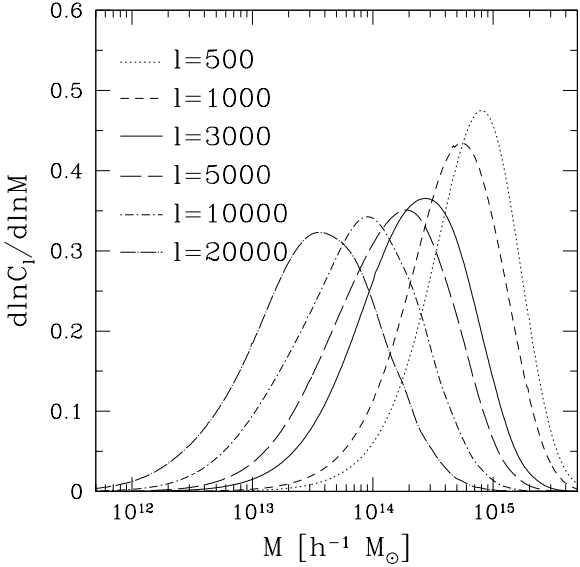}
  \end{center}
\caption[]
 {Mass distribution of $C_l$.
 We plot $d\ln C_l/d\ln M$, equation~(\ref{eq:dlncldlnM}), for a given $l$.
 From right to left it is shown $l=500$, 1000, 3000, 5000, 10000,
 and 20000, as indicated in the figure.
 }
\label{fig:dcdM}
\end{figure}

Figure~\ref{fig:dcdM} shows the mass distribution of $C_l$ for a given $l$, 
\begin{equation}
 \label{eq:dlncldlnM}
 \frac{d\ln C_l}{d\ln M}\equiv 
 \frac{M \int dz \frac{dV}{dz}
 \frac{dn(M,z)}{dM}
	      \left|\tilde{y}_l(M,z)\right|^2}
 {\int dz \frac{dV}{dz}
 \int dM \frac{dn(M,z)}{dM}
	      \left|\tilde{y}_l(M,z)\right|^2}.
\end{equation}
We find that massive haloes between $10^{14}~h^{-1}~M_{\sun}$
and $10^{15}~h^{-1}~M_{\sun}$ dominate $C_l$ at $l=3000$ with a peak 
at $3\times 10^{14}~h^{-1}~M_{\sun}$. 
For $l=1000$ the peak is at $5\times 10^{14}~h^{-1}~M_{\sun}$, while for 
$l=10000$ at $10^{14}~h^{-1}~M_{\sun}$. 
The large-angular-scale $C_l$ is thus sensitive to
the presence of very massive and rare haloes, which
explains why the sampling variance is so large on large scales. 
Much smaller angular scales ($<1'$ or $l>10000$) receive significant 
contribution from smaller mass halos, where uncertain physics of gas cooling, 
heating, or star formation makes predictions unreliable; however, 
for the angular scales of $2'-4'$ the dominant 
contribution is from massive ($10^{14}-10^{15}~h^{-1}~M_{\sun}$) clusters 
for which these uncertainties are small and which are still so abundant 
that the sampling variance is small. 
Fortunately, this is precisely the range targeted by the next generation 
CMB experiments.

\section{Covariance matrix of the SZ angular power spectrum}
\label{sec:error}

To determine the cosmological parameters from the power spectrum 
we need to know its covariance matrix.
If fluctuations are Gaussian, then we can calculate the variance from 
the spectrum itself.
If fluctuations are non-Gaussian, then the covariance matrix
is no longer diagonal. 
In this case we need to know not only $C_l$, but also specific 
configurations of the angular trispectrum, the harmonic transform of the 
four-point correlation function, to estimate the covariance matrix.

The SZ fluctuations observed in the hydrodynamic simulations are very 
non-Gaussian \cite{2001PhRvD..630f619S,2002astro.ph..1375Z,astro-ph/0107023}.
The scatter among the $C_l$ estimates is large, especially on 
large scales, and the estimates are strongly correlated. 
We must thus calculate the full covariance matrix, taking into
account the non-Gaussianity. 
Here we will use both the halo approach \cite{2001PhRvD..64f3514C}, and 
the hydrodynamic simulations \cite{2001PhRvD..630f619S}, to develop a 
reliable method for covariance matrix calculation.

The covariance matrix of $C_l$, $M_{ll'}$, is given by 
\cite{2001PhRvD..64f3514C}
\begin{eqnarray}
 \nonumber
 M_{ll'}&\equiv&
  \left<\left(C^{\rm obs}_l-C_l\right)
   \left(C^{\rm obs}_{l'}-C_{l'}\right)\right>\\
 \label{eq:cov}
  &=&
  f_{\rm sky}^{-1}
  \left[\frac{2\left(C_l\right)^2}{2l+1}\delta_{ll'}
 +\frac{T_{ll'}}{4\pi}\right],
\end{eqnarray}
where the angular bracket denotes the ensemble average,
$f_{\rm sky}$ is a fraction of the sky covered by an experiment
($f_{\rm sky}=1$ for all sky),
$C_l^{\rm obs}$ the observationally measured power spectrum,
and $C_l$ the theoretically predicted power spectrum (equation~\ref{eq:cl}).
Note that $\left<C_l^{\rm obs}\right>=C_l$.
The second term in the r.h.s. represents the trispectrum contribution.

\subsection{Halo approach to the angular trispectrum}

In this section we calculate $M_{ll'}$, equation~(\ref{eq:cov}), 
using the halo approach, and compare it with the hydrodynamic simulations.
An angular-trispectrum configuration is represented by a quadrilateral, 
and is characterized by four sides and one diagonal \cite{2001PhRvD..64h3005H}.
Among the trispectrum configurations, those which constitute two lines 
whose lengths are $l$ and $l'$, respectively, and have zero diagonal, 
determine the power-spectrum covariance matrix. 
We denote these as $T_{ll'}$ (equation~\ref{eq:cov}).

\begin{figure}
  \begin{center}
    \leavevmode\epsfxsize=8.4cm \epsfbox{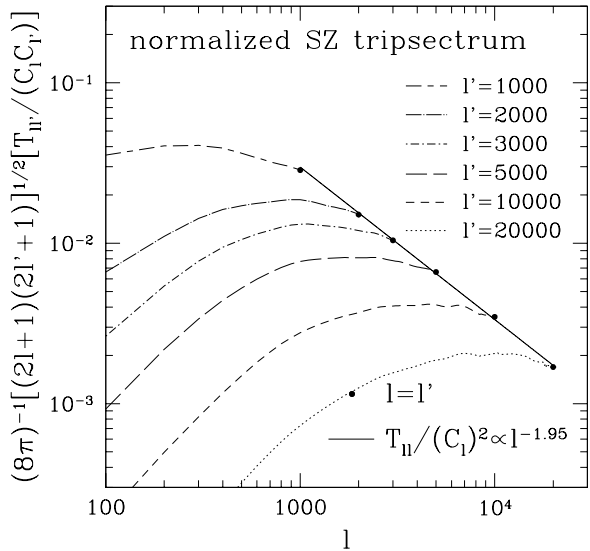}
  \end{center}
\caption[]
 {The SZ angular trispectrum, $T_{ll'}$, multiplied by 
 $\sqrt{(2l+1)(2l'+1)}/C_lC_{l'}$.
 From top to bottom it is shown 
 $l'=1000$, 2000, 3000, 5000, 10000, and 20000, as indicated in the figure.
 The solid line shows $\propto l^{-0.95}$, which means
 that the ratio of the non-Gaussian error to the Gaussian error
 decreases with $l$ as $l^{-0.95}$, and  
 $T_{ll}$ scales as $\propto l^{-1.95}(C_l)^2$.
 The plotted quantity reduces to the correlation coefficient 
 of $C_l$ in the limit of $T_{ll}\ll (C_l)^2$.}
\label{fig:trispectrum}
\end{figure}

Using the halo approach one obtains for the Poisson term which dominates
the trispectrum \cite{2001PhRvD..64f3514C},
\begin{eqnarray}
 \nonumber
 T_{ll'}&=& g_\nu^4 \int_0^{z_{\rm max}} dz \frac{dV}{dz}
              \int_{M_{\rm min}}^{M_{\rm max}} dM \frac{dn(M,z)}{dM}\\
 \label{eq:tl}
	& &\times \left|\tilde{y}_l(M,z)\right|^2
	\left|\tilde{y}_{l'}(M,z)\right|^2,
\end{eqnarray}
where all the quantities have the same meaning
as in equation~(\ref{eq:cl}).
It follows from equation~(\ref{eq:cov}) that 
$(4\pi f_{\rm sky})^{-1}T_{ll'}$ determines the non-diagonal terms of 
$M_{ll'}$.
Figure~\ref{fig:trispectrum} shows $T_{ll'}$ divided by $C_lC_{l'}$.
To make it represent a ratio of the ``non-Gaussian error''
(the second term in $M_{ll'}$ of equation~\ref{eq:cov}) 
to the ``Gaussian error'' (the first term) we have plotted 
$T_{ll'}/(C_lC_{l'})$ multiplied by $\sqrt{(2l+1)(2l'+1)}/8\pi$, which is 
this ratio when $l=l'$.
We find that the ratio decreases with $l$ as $l^{-0.95}$, or 
equivalently, $T_{ll}\propto l^{-1.95}(C_l)^2$, for $l\ga 10^3$.

Figure~\ref{fig:trispectrum} also shows how $M_{ll'}$ 
scales with $l$ for a given $l'$.
In the limit of $T_{ll}\ll (C_l)^2$, the plotted quantity in 
figure~\ref{fig:trispectrum} reduces to the correlation coefficient 
of $C_l$, $r_{ll'}$, given by
\begin{eqnarray}
 \nonumber
  r_{ll'}
  &\equiv& \frac{M_{ll'}}{\sqrt{M_{ll}M_{l'l'}}}\\
 \label{eq:ratio}
  &\approx&
  \frac{\sqrt{(2l+1)(2l'+1)}}{8\pi}
  \frac{T_{ll'}}{C_lC_{l'}}
  \qquad
  \left(T_{ll}\ll C_l^2\right).
\end{eqnarray}
As $T_{ll'}/(C_lC_{l'})$ varies slowly with $l$ for a 
given $l'$, $C_l$ and $C_{l'}$ remain highly correlated
even for a large value of $\left|l-l'\right|$ \cite{2001PhRvD..64f3514C}. 
Because of this reason binning does not reduce the error bars as much as 
it would for Gaussian fluctuations.

\subsection{Comparison with the simulated error bars}

To compare the predicted error bars of $C_l$ with the simulated ones 
one must take into account binning or band averaging that 
the simulations use to produce $C_l$.
If the binned $C_l$ at a given $l$ is the average of $C_l$
between $l-\Delta l/2 \leq l \leq l+\Delta l/2$, then the covariance matrix is
\begin{equation}
 \label{eq:cov_bin}
 M_{ll'}\approx
  f_{\rm sky}^{-1}
  \left[\frac{2\left(C_l\right)^2}{(2l+1)\Delta l}\delta_{ll'}
 +\frac{T_{ll'}}{4\pi}\right].
\end{equation}
Here, we have assumed that $\Delta l\ll l$.
Since both $C_l$ and $T_{ll'}$ are sufficiently smooth,
the binning does not affect the values of $C_l$ and $T_{ll'}$
as long as $\Delta l\ll l$.
We find that $\Delta l/l< 0.1$ suffices for this approximation to hold.

\begin{figure}
  \begin{center}
    \leavevmode\epsfxsize=8.4cm \epsfbox{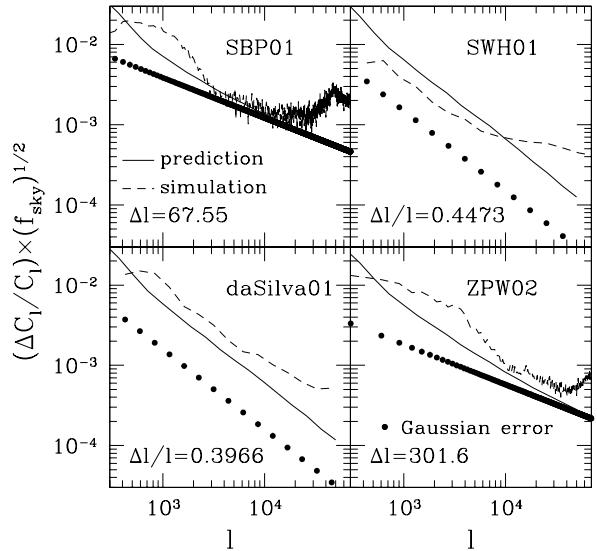}
  \end{center}
\caption[]
 {Predicted fractional r.m.s. errors of $C_l$, 
 $\Delta C_l/C_l\equiv \sqrt{M_{ll}}/C_l$, in comparison with the simulated 
 errors (see table~\ref{tab:simulations} for the meaning of the labels 
 shown in each panel, and for the cosmological parameters used).
 We have multiplied the simulated errors by the square-root of 
 a fraction of the sky covered by the simulations, $f_{\rm sky}$.
 The solid lines show the predicted fractional errors, while the dashed
 lines show the simulated ones.
 The filled circles show Gaussian approximation to the fractional errors,
 $\simeq (l\Delta l)^{-1/2}$ for $l\gg 1$, where $\Delta l$ is the 
 binning size shown in each panel.
 The predicted errors agree with the simulated ones within a factor of
 two, providing a significant improvement over the Gaussian approximation.}
\label{fig:err}
\end{figure}

Figure~\ref{fig:err} compares the predicted fractional errors
of $C_l$, $\Delta C_l/C_l\equiv \sqrt{M_{ll}}/C_l$, with the simulated ones.
We find that the predictions agree with the simulations within a factor 
of two, similar to the agreement we have found for the spectrum itself.
There is no clear trend in the residuals and we both ovepredict and 
underpredict the variance compared to the simulations. 
Since the variance is very sensitive to the sampling errors, it is possible 
that some of the discrepancy is caused by insufficient sampling in the 
simulations.
Figure~\ref{fig:err} also plots Gaussian approximation to the errors,
showing that our predictions give a significant improvement over the 
Gaussian approximation.
We have also compared the predicted correlation coefficient, $r_{ll'}$
(equation~\ref{eq:ratio}), with the simulations of 
\shortciteN{2001PhRvD..630f619S}, 
finding a good agreement between the two and confirming that for high $l$
the neighboring bins in $C_l$ are strongly correlated if the binning is 
sufficiently broad for the trispectrum term to dominate. 
Proper modeling of cross-correlations between the bins is essential for 
the parameter-error estimation and for the proper treatment of 
the statistical significance of results. 

\section{Determination of power-spectrum amplitude}
\label{sec:observations}

In this section, we analyze how well we can determine 
$\sigma_8$ by measuring the SZ angular power spectrum with realistic 
observations, and apply it to the recent CBI results \cite{astro-ph/0205384}.

We perform a least-square fitting to $C_l$ by calculating $\chi^2$,
\begin{equation}
 \label{eq:chisq}
  \chi^2  \equiv
  \sum_{l\le l'}
  \left(\hat{C}_l-C_l\right)\left(M^{-1}\right)_{ll'}
  \left(\hat{C}_{l'}-C_{l'}\right),
\end{equation}
where $\hat{C}_l$ is a reference power spectrum which uses
the fiducial cosmological parameters, or the actual CBI data.
We determine the confidence levels of the parameter estimation by 
calculating $\Delta\chi^2$.

We vary $\Omega_{\rm m}$ and $\sigma_8$, i.e.,
$\chi^2=\chi^2\left(\Omega_{\rm m}, \sigma_8\right)$.
We fix the other parameters at the fiducial values, 
$\Omega_{\Lambda}=1-\Omega_{\rm m}$, $\Omega_{\rm b}=0.05$, 
$h=0.7$, $w=-1.0$, and $n=1.0$.
For a given set of cosmological parameters, we calculate 
$C_l$ and $M_{ll'}$ and then obtain $\chi^2$.
By approximating probability distribution function of 
$C_l$ with a Gaussian  we have $\Delta\chi^2=2.30$, 6.17, 
and 11.8 for 68.3\%, 95.4\%, and 99.7\% confidence levels (C.L.), respectively.

First, we assume that experiments are sampling-variance limited
for $2000<l<3000$, or $2000<l<5000$; thus, we consider 
a few arcminutes angular resolution and ignore instrumental noise.
To save computational time we do not evaluate equation~(\ref{eq:chisq})
at every $l$, but at 10 points for $2000<l<3000$, or
20 points for $2000<l<5000$.
This choice gives $\Delta l=111$ and 158, respectively, in 
equation~(\ref{eq:cov_bin}).
Both $C_l$ and $T_{ll'}$ are smooth so that this is a good approximation.

For the sky coverage we consider 
$1~{\rm deg^2}$ or $10~{\rm deg^2}$ survey of the sky.
These parameters are characteristic for the on-going or forthcoming
experiments: CBI \cite{2001ApJ...549L...1P} 
for $2000<l<3000$ and $1~{\rm deg^2}$,
SZA \cite{2001ApJ...558..515H} for $2000<l<3000$ and $10~{\rm deg^2}$,  
AMIBA \cite{2002astro.ph..1375Z} or ACT (Lyman Page, private communication)
for $2000<l<5000$ and $10~{\rm deg^2}$.
AMIBA and ACT experiments are actually going to survey about 
$100~{\rm deg^2}$ of the sky, so that the actual constraints to be obtained
from these experiments would be better than we show here.

\begin{figure}
  \begin{center}
    \leavevmode\epsfxsize=8.4cm \epsfbox{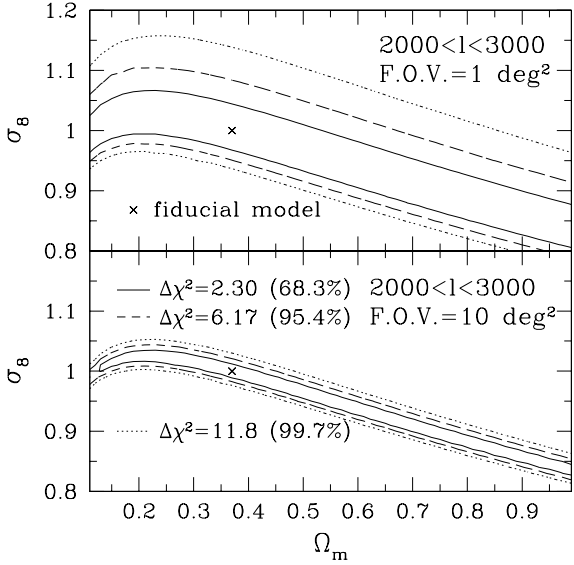}
  \end{center}
\caption[]
 {Contours showing how well we can determine $\sigma_8$ against
 $\Omega_{\rm m}$ by measuring the SZ angular power spectrum with 
 realistic observations.
 We consider those observations which measure $C_l$
 between $2000<l<3000$, and survey $1~{\rm deg^2}$ (top panel), or 
 $10~{\rm deg^2}$ (bottom panel) of the sky.
 The top panel is similar to the CBI experiment and the bottom panel
 to the SZA experiment. 
 The solid line shows $\Delta\chi^2=2.30$ (68.3\% C.L.),
 the dashed line $\Delta\chi^2=6.17$ (95.4\% C.L.), and
 the dotted line $\Delta\chi^2=11.8$ (99.7\% C.L.).
 The cross marks the fiducial cosmological model.
 We find that the contour is very sensitive to $\sigma_8$, yet almost 
 independent of $\Omega_{\rm m}$
 at around 
 $\Omega_{\rm m}\sim 0.2$ (see also
 section~\ref{sec:simulations}).
 We find that a CBI-type observation should be able to determine 
 $\sigma_8$ to within 10\% at $2\sigma$, while a SZA-type
 to within 5\% at the same level.}
\label{fig:l3000}
\end{figure}

Figure~{\ref{fig:l3000}} shows $\chi^2$ on a 
$\Omega_{\rm m}-\sigma_8$ plane for the $2000<l<3000$ region.
The solid line shows $\Delta\chi^2=2.30$ (68.3\% C.L.),
the dashed line $\Delta\chi^2=6.17$ (95.4\% C.L.), and
the dotted line $\Delta\chi^2=11.8$ (99.7\% C.L.).
The top panel uses $1~{\rm deg^2}$ of the sky, close to the CBI
experiment (if it is sampling-variance limited), while the bottom panel 
uses $10~{\rm deg^2}$ of the sky, close to the SZA experiment. 
We find that the contours are very sensitive to $\sigma_8$ and almost
independent of $\Omega_{\rm m}$ for $0.15<\Omega_{\rm m}<0.4$ 
\cite{1999ApJ...526L...1K}.
This is consistent with what we have previously shown in
section~\ref{sec:simulations} (see equation~\ref{eq:fit}).
While for one degree sky coverage the sampling-variance error is similar 
to the systematic (theoretical) error (10\% at the 95\% confidence level), 
a 10-times larger survey like SZA with the same angular resolution should be
able to determine $\sigma_8$ to within 5\% at the same level, so at that 
point the errors will be dominated by the theoretical uncertainty
rather than by statistics.

\begin{figure}
  \begin{center}
    \leavevmode\epsfxsize=8.4cm \epsfbox{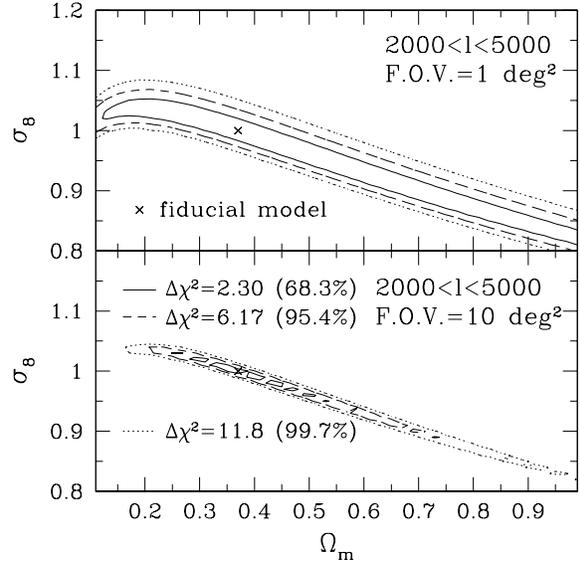}
  \end{center}
\caption[]
 {The same figure as figure~\ref{fig:l3000}, but for $2000<l<5000$.
 We find that statistical errors on $\sigma_8$ are very small
 for a $10~{\rm deg^2}$ survey, showing that ACT or AMIBA
 experiments which survey about $100~{\rm deg^2}$ of the sky 
 determine $\sigma_8$ with theoretical-uncertainty-limited
 accuracy ($\sim 10\%$).
 A $1~{\rm deg^2}$ survey shown in the top panel is
 sufficient enough to determine $\sigma_8$ with that accuracy.}
\label{fig:l5000}
\end{figure}

Figure~\ref{fig:l5000} shows constraints on a $\Omega_{\rm m}-\sigma_8$
plane which would be obtained with ACT or AMIBA, 
whose angular resolution is $1'-2'$. 
The figure uses $2000<l<5000$ with $1~{\rm deg^2}$ (top panel) or 
$10~{\rm deg^2}$ (bottom panel) sky coverage.
We find that the statistical error on $\sigma_8$ is extremely small
for a $10~{\rm deg^2}$ survey. Surveys such as ACT or AMIBA
covering about $100~{\rm deg^2}$ of the sky 
will determine $\sigma_8$ with accuracy limited by the theoretical uncertainty.
This is currently at a $\sim 10\%$ level and obviously 
more theoretical work should improve it in order for the power of the 
upcoming CMB experiments to be maximally exploited.
It is of course still useful to cover such a large area of the sky,
since this will allow us to use a bootstrap error determination without 
relying on the simulations. 

\begin{figure} 
  \begin{center} 
    \leavevmode\epsfxsize=8.4cm \epsfbox{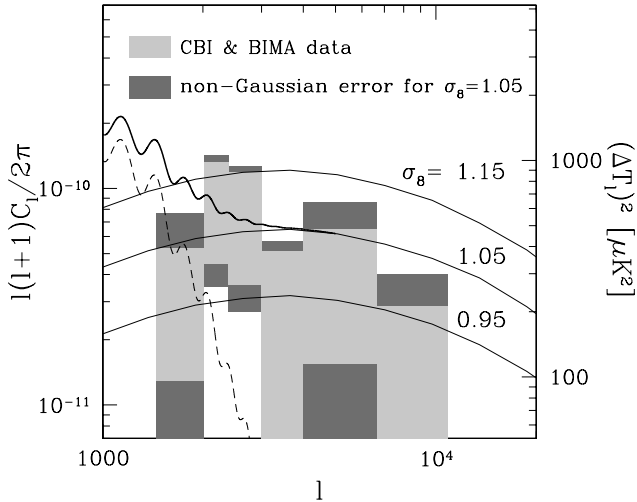} 
  \end{center} 
\caption[] 
 {The CBI \cite{astro-ph/0205384} and BIMA \cite{2002astro.ph..6012D} data
  with Gaussian errors (light-gray shaded area), the predicted SZ angular 
  power spectra for $\sigma_8=0.95$, 1.05, and 1.15 (thin solid lines), 
  the primary CMB anisotropy for the fiducial cosmological model 
  (dashed line), and the sum of the two for $\sigma_8=1.05$ (thick solid line).
  The dark-shaded area gives non-Gaussian errors predicted for
  $\sigma_8=1.05$.}
\label{fig:cbibima} 
\end{figure}

Recently, CBI and BIMA announced detections of CMB power at
$l \sim 2000-6000$, which at face value are inconsistent with the 
primary CMB alone \cite{astro-ph/0205384,2002astro.ph..6012D}. 
Figure~\ref{fig:cbibima} compares the CBI and BIMA data
(light-gray shaded area, which assumes Gaussian errors) to the predicted SZ 
angular power spectra for $\sigma_8=0.95$, 1.05, and 1.15 
(thin solid lines). 
We also show the primary CMB anisotropy for the fiducial 
cosmological model (dashed line), and the sum of the two for 
$\sigma_8=1.05$ (thick solid line).
The predicted non-Gaussian errors are shown for $\sigma_8=1.05$
(dark-gray shaded area).
We have included the instrumental-noise power spectrum in the non-Gaussian
errors by using the noise power spectrum of
$l(l+1)C_l/(2\pi)=500$, 1000, 1500, and $3000~\mu{\rm K}^2$ 
for the CBI bins, and 720 and 2000~$\mu{\rm K}^2$ for the BIMA bins.
We have assumed the sky coverage of $1~{\rm deg^2}$ for CBI and
$0.1~{\rm deg}^2$ for BIMA.
From the figure one can see that the data favor $\sigma_8\sim 1$
models, and the predicted power spectra can explain the detected excess power 
reasonably well.

\begin{figure} 
  \begin{center} 
    \leavevmode\epsfxsize=8.4cm \epsfbox{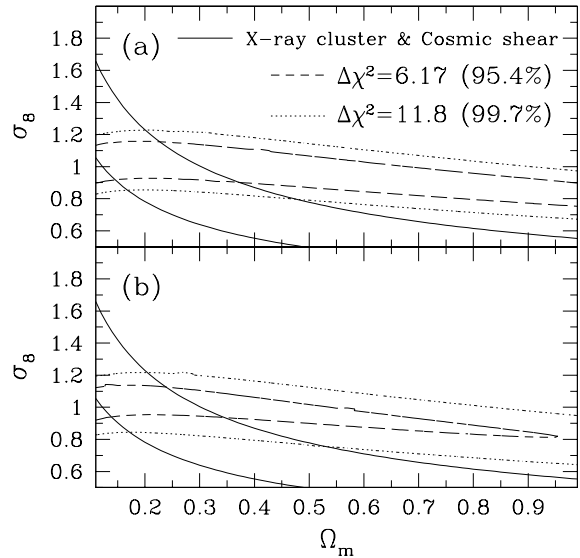} 
  \end{center} 
\caption[] 
 {The CBI and BIMA constraints on the $\Omega_{\rm m}-\sigma_8$ plane.
 The top panel (a) assumes that all the data shown in figure~\ref{fig:cbibima}
 are entirely due to the SZ effect. 
 The dashed line shows $\Delta\chi^2=6.17$ (95.4\% C.L.) while
 the dotted line $\Delta\chi^2=11.8$ (99.7\% C.L.).
 The cluster abundance and cosmic-shear constraints 
 are shown in the figure as $\sigma_8\Omega_{\rm m}^{0.5}=0.45\pm 0.1$, which 
 roughly summarizes the current uncertainty from these methods.
 The bottom panel (b) assumes that the first bin of the CBI data is affected
 by the primary CMB anisotropy.
 In both cases, we find $\sigma_8(\Omega_{\rm b}h/0.035)^{2/7}=1.04\pm 0.12$ 
 at the 95.4\% confidence level for $0.1<\Omega_{\rm m}<0.5$.} 
\label{fig:cbibimafit} 
\end{figure}

We fitted the data to the predicted SZ angular power spectrum using
the four bins shown in figure~\ref{fig:cbibima} and computing
$\chi^2$ from equation~(\ref{eq:chisq}). 
We compute the covariance matrix using equation~(\ref{eq:cov_bin}) with
the bin-size of $\Delta l=565$, 378, 612, 1000, 2870, and 4150 for each bin.
For example, we have obtained r.m.s. error of 
327, 398, 373, 425, 370, and 299~$\mu~{\rm K}^2$ at each bin for 
$\sigma_8=1.05$.
The cross-correlation between the bins is most significant for the first 
bin, for which the correlation coefficients with the other 3 CBI bins are 
0.46, 0.41, and 0.29.
The remaining 3 coefficients are below 0.3. 
Figure \ref{fig:cbibimafit} shows confidence-level contours on the 
$\Omega_{\rm m}-\sigma_8$ plane for the CBI and BIMA data.
The top panel (a) assumes that the detection of the excess power in the 
all bins is entirely due to the SZ effect 
(i.e., we ignore the primary CMB anisotropy).
We find $\sigma_8(\Omega_{\rm b}h/0.035)^{2/7}=1.04\pm 0.12$ at 95\% 
confidence limit for $0.1<\Omega_{\rm m}<0.5$.

The cluster abundance and cosmic-shear constraints are also shown in the 
figure as $\sigma_8\Omega_{\rm m}^{0.5}=0.45\pm 0.1$, which roughly 
summarizes the current (systematic dominated) uncertainty from these methods.
Combining the two can break the degeneracy between 
$\sigma_8$ and $\Omega_{\rm m}$ and is leading to $\sigma_8 \sim 1$, 
$\Omega_{\rm m} \sim 0.2$. 
One should be cautious not to overinterpret this conclusion, since 
systematic effects in both are still very 
large and could affect the parameter determination.

The bottom panel (b) assumes that the $l \sim 1700$ bin is affected by 
the primary CMB anisotropy, while the other bins are entirely
due to the SZ effect. 
We assume that the primary CMB is $400~{\mu{\rm K}^2}$ at the first bin, 
roughly consistent with the best-fitting theoretical curve, and is 
negligible in the other three bins.
In principle we should change the primary CMB as well as we change 
the cosmological parameters; however, for simplicity we just use 
$400~{\mu{\rm K}^2}$ regardless of the cosmological parameters. 
We find a similar constraint on the $\Omega_{\rm m}-\sigma_8$ plane 
to the panel (a).
In general, the CBI and BIMA data are consistent with detection of the 
SZ fluctuations, and imply 
$\sigma_8(\Omega_{\rm b}h/0.035)^{2/7}=1.04\pm 0.12$, but
a somewhat larger survey is needed to confirm this interpretation. 
The good news is that with a survey just a few times the 
survey area of CBI one should be able to determine $\sigma_8$ with an 
accuracy comparable to or better than any other current observations and with 
a different, and perhaps smaller, systematic uncertainty. 
    
\section{Uncertainties in the predictions}\label{sec:uncertainty}

The predictions are very sensitive to the halo mass
function, so we investigate
how the prediction changes if using different formulae in the literature.
We compare four different predictions for $C_l$ computed with
four different formulae for the mass function:
(JB3) Jenkins et al.'s $\tau$CDM mass function that uses the SO mass with
$\delta=180\Omega_{\rm m}(z)$ \cite{2001MNRAS.321..372J},
(J9) Jenkins et al.'s universal mass function that uses the FOF mass
(equation~9 of \citeNP{2001MNRAS.321..372J}) and includes smoothing which 
may systematically increase the mass function in the exponential tail 
 \cite{2002astro.ph..3169H}, 
(HV) Evrard et al.'s Hubble-volume mass function which uses the SO mass with
$\delta=200$ \cite{2002ApJ...573....7E}, and
(PS) Press \& Schechter's mass function which uses the virial mass
\cite{1974ApJ...187..425P} 
(this one is of interest just for historical comparison). 
As \citeN{2001MNRAS.321..372J} note that the FOF mass they use
corresponds to the SO mass with $\delta=180\Omega_{\rm m}(z)$, we
use the SO mass to calculate the mass function of (J9).

\begin{figure}
  \begin{center}
    \leavevmode\epsfxsize=8.4cm \epsfbox{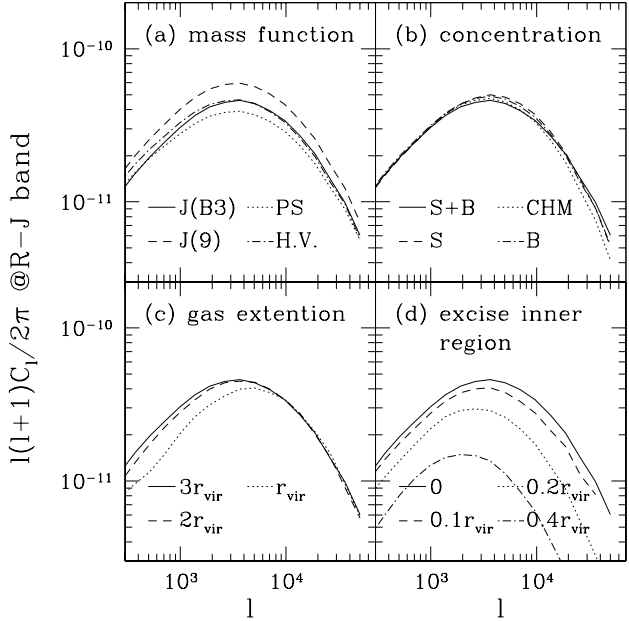}
  \end{center}
\caption[]
 {Theoretical uncertainty in the SZ angular power spectrum.
 {\bf (a)} Uncertainty in the dark-matter-halo mass function.
 The solid line shows the prediction with Jenkins et al.'s $\tau$CDM 
 mass function 
 (equation~\ref{eq:dndM_jenkinsB3}),
 the dashed line Jenkins et al.'s universal 
 mass function (equation~9 of \citeN{2001MNRAS.321..372J}), and
 the dot-dashed line Evrard et al.'s Hubble-volume mass function 
 \cite{2002ApJ...573....7E}.
 For comparison, the dotted line plots the prediction with
 Press \& Schechter's mass function (Press \& Schechter 1974).
 {\bf (b)} Uncertainty in the concentration parameter.
 {\bf (c)} Uncertainty in the hot-gas extension in haloes.
 The solid line shows $C_l$ with the hot-gas extension
 cut out at 3 times the virial radius,
 the dashed line at 2 times the virial radius,
 and the dotted line at the virial radius. 
 {\bf (d)} Uncertainty in the inner gas-pressure profile.
 The solid line shows $C_l$ with no excision, 
 the dashed line shows the one with the region interior to 
 10\% of the virial radius excised,
 the dotted line 20\% of the virial radius,
 and the dot-dashed line 40\% of the virial radius. The dominant contribution to 
the spectrum is from $20-40\%$ of the virial radius}
\label{fig:systematics}
\end{figure}

The top-left panel~(a) of figure~\ref{fig:systematics} compares
$C_l$ computed with the four different mass functions.
We find a very good agreement between (JB3) and (HV), while (J9) predicts
higher amplitude.
\citeN{2002astro.ph..3169H} have also found that the mass function of (J9)
predicts too many high-mass haloes to agree with their $N$-body
simulations, and concluded that the smoothing used for deriving
(J9) causes the high-mass halo abundance to increase anomalously.
\citeN{2001MNRAS.321..372J} have explained this in their paper: 
when deriving (J9), they are mainly interested in small-mass haloes 
for which the smoothing is less significant.
Press \& Schechter's mass function predicts lower amplitude
and underestimates the abundance of 
high-mass haloes compared with $N$-body simulations
\cite{2001MNRAS.321..372J}.
As the SZ effect is dominated by the high-mass haloes, we should not 
use (J9) and (PS), but (JB3) or (HV).

Since there is no single expression for the concentration
parameter $c$ (see equation~\ref{eq:c} for definition) we need to
quantify how much uncertainty in $C_l$ is due to the uncertainty 
in the concentration parameter.
To quantify this, we use four different formulae for the concentration
parameter:
(S+B) power-spectrum-determined
concentration parameter at $z=0$ \cite{2000MNRAS.318..203S} but 
evolved by $(1+z)^{-1}$, for which evolution 
has been measured in $N$-body simulations \cite{2001MNRAS.321..559B}, 
(S) power-spectrum determination with the non-linear mass evaluated at $z$,
(B) halo determination of the concentration parameter 
\cite{2001MNRAS.321..559B} and
(CHM) a different power-spectrum-determined concentration parameter 
\cite{2000ApJ...535L...9C}.
The top-right panel~(b) of figure~\ref{fig:systematics} 
shows the predictions computed with the four different concentration 
parameters.
We find that the uncertainty in the concentration parameter 
has little effect on $C_l$.

How sensitive are the results to the outer radius of the gas 
profile (shock radius), where the gas temperature rapidly drops
to IGM values?  
Previous work on a gas profile has assumed that
the hot gas in haloes extends only up to the virial radius, but numerical 
simulations suggest a smooth decrease in temperature at least out to twice the 
virial radius 
\cite{1998ApJ...495...80B,1998ApJ...503..569E,1999ApJ...525..554F}.
The bottom-left panel~(c) of figure~\ref{fig:systematics}
shows the predicted $C_l$ with the hot-gas extension cut out
at 1, 2, or 3 times the virial radius.
While a cut-off at the virial radius suppresses $C_l$ 
at $l<5000$, there is little difference for the cut-off at 
$2 r_{\rm vir}$ and beyond.
One should thus include the hot-gas extension
beyond the virial radius; otherwise, one
underestimates the power spectrum
on large angular scales, but the actual position 
of the cut-off is not very important since the power 
spectrum converges at $2 r_{\rm vir}$.

The gas density profile in the inner region displays a core
\cite{1998ApJ...495...80B,1998ApJ...503..569E,1999ApJ...525..554F}, 
which is well reproduced by our model \cite{2001MNRAS.327.1353K}. 
On the other hand, recent X-ray observations have observed a
decrease in temperature toward the centre within 5\% of the virial radius
\cite{2000ApJ...534L.135M,2001MNRAS.324..877A,2001MNRAS.324..842A,2001A&A...365L..67A,2001A&A...365L..80A,2001A&A...365L..99K}, 
which our polytropic model does not account for.
The bottom-right panel~(d) of figure~\ref{fig:systematics}
shows the predicted $C_l$ with the inner region within
10\%, 20\%, or 40\% of the virial radius excised. 
We find that the region interior to 10\% of the virial radius
has a negligible contribution to $C_l$, so the 
pressure profile in the cluster core has no effect on $C_l$.
From this figure we find that the dominant contribution to $C_l$
comes from around $20-40\%$ of the virial radius, 
roughly corresponding to $r_{2500}$, which is just in the range 
of {\it Chandra} satellite for medium redshift clusters
\cite{2000ApJ...534L.135M,2001MNRAS.324..877A,2001MNRAS.324..842A} 
and well within {\it XMM/Newton} or {\it BeppoSax} 
\cite{2001A&A...365L..67A,2001A&A...365L..80A,2001A&A...365L..99K}. 
Recent observations confirm that the gas profile 
agrees well with the dark-matter profile in this region 
\cite{2002astro.ph..5007A}, confirming our basic assumption. 
We note that the temperature gradient is still small over this range, 
so most of the pressure gradient is caused by the gas-density profile.

The effect of non-adiabatic physics has been investigated with 
hydrodynamic simulations. In \citeN{2001ApJ...561L..15D} it has been 
claimed that radiative cooling reduces $C_l$ by modest amount, 
$20-40\%$, on all angular scales.
This is presumably due to a decrease in the gas fraction, since some fraction 
of gas is being transformed into stars.
Despite their extreme cooling treatment (giving
too much cooled material to match the observations)
the effect of radiative cooling is modest
(note that 40\% change in $C_l$ gives a 5\% change in $\sigma_8$).
Preheating may lead to a larger effect, as it
may alter the gas structure not only in the inner region, but
also at larger radii, especially for smaller-mass haloes.
Some fraction of the gas may be expelled from the cluster, which 
would suppress the SZ fluctuations on small scales. 
At the same time, preheating increases the gas temperature, which would 
increase the spectrum.
However, care must be taken not to violate the FIRAS constraints on 
the mean $y$ parameter and it is unlikely that this could significantly 
enhance the SZ power spectrum without violating these limits.
Several authors have noticed significant effects of preheating on $C_l$
\cite{2001ApJ...549..681S,2001ApJ...558..515H,2001ApJ...561L..15D}, 
although current treatments are still over-simplified. 
One must be careful in analytic models to make them
self-consistent, as one cannot retain gas in hydrostatic equilibrium within
external dark-matter potential using 
a large gas core and isothermal gas. 
When hydrostatic equilibrium is imposed the pressure profile 
is less affected than the density or temperature profiles. 
It is likely that any preheating will have only a modest effect on the 
SZ power spectrum. 
Recently, \citeN{astro-ph/0107023} have used hydrodynamic simulations to 
show that gas cooling, energy feedback, and star formation affect $C_l$ 
by no more than a factor of two, confirming our conclusion.

\section{Conclusions}
\label{sec:conclusions}

In this paper we present a refined analytic model for the angular power 
spectrum of the SZ effect.
In contrast to previous modeling 
\cite{1999ApJ...515..465A,1999ApJ...526L...1K,2000PhRvD..62j3506C,2000ApJ...537..542M,2001ApJ...558..515H,2001ApJ...549...18Z}, our model 
has no free parameters  and treats the gas in a cluster dark-matter 
potential self-consistently, 
using the universal gas-density and temperature
profile \cite{2001MNRAS.327.1353K}, which fits the observed and the 
simulated properties of clusters of galaxies in the range of interest to 
the SZ observations (outside 5\% of the virial radius).

We compare our predictions for the spectrum and the errors with 
the hydrodynamic SZ simulations and find a good agreement.
The deviations between our model and the simulations are comparable to the 
deviations among the simulations themselves and are particularly small in the 
observationally relevant range around $l \sim 3000$. 
Some of the discrepancies are due to either
poor resolution in the simulations or sampling variance, while 
the remaining difference could be due to the simplifying assumptions of gas physics 
in our model, such as hydrostatic equilibrium, spherical symmetry, merging and 
substructure, which 
may be inaccurate in particular at higher redshifts when the clusters
are still forming. 
The disagreement is more important on very small angular scales 
($l>5000$), while the accuracy of our predictions in the range of 
$l\sim 2000-5000$ is better than a factor of two, which translates to
less than 10\% systematic uncertainty in the amplitude of fluctuations
$\sigma_8$. 
We also compare analytic predictions for the power-spectrum covariance 
matrix to the simulations and find similarly good agreement.

We investigate the dependence of $C_l$ on various cosmological parameters,
finding that over the range of interest $\sigma_8$ and $\Omega_{\rm b}h$
determine the amplitude of $C_l$ almost entirely,
as has been pointed out by several authors before as well
\shortcite{1999ApJ...526L...1K,2001PhRvD..630f619S,2001ApJ...549...18Z,2002astro.ph..1375Z}.
It is particularly important that the matter density of the universe 
does not affect the spectrum as much as $\sigma_8$ does,
in agreement with \citeN{1999ApJ...526L...1K}.
This differs from the local-cluster-abundance studies, where the constraint 
is usually on $\sigma_8\Omega_{\rm m}^{0.5}$.

The dominant contribution to $C_l$ comes from 
massive clusters ($M>10^{14}~h^{-1}~M_{\sun}$ for $l<5000$)
at moderate ($z \sim 1$) redshift. 
Within the cluster, the dominant contribution comes from around 
$0.2-0.4r_{\rm vir}$, suggesting that the physical processes taking place
in the cluster core such as cooling, heating or heat conduction have 
little effect on the SZ power spectrum. 
We assume that the gas fraction is close to the cosmic mean baryon fraction, 
which is reasonable for massive clusters and has observational support in 
low stellar-mass-to-cluster-mass ratios in massive clusters 
(e.g., \citeNP{1998ApJ...503..518F}), as well as 
in studies of the gas fraction as a function of radius 
(e.g., \citeNP{2002astro.ph..5007A}).
While more exotic mechanisms such as preheating could affect $C_l$ on all 
scales (by increasing the temperature and/or reducing the gas clumping), 
they are constrained by the FIRAS limits and are unlikely to make more than a 
factor of two
change in $C_l$ (comparable to a 10\% change in $\sigma_8$).

Using analytic predictions for the spectrum and its covariance matrix
we have performed a likelihood analysis to estimate how well
we can measure $\sigma_8$ against $\Omega_{\rm m}$ with realistic SZ
observations.
The likelihood is very sensitive to $\sigma_8$
almost independent of $\Omega_{\rm m}$ (figure~\ref{fig:l3000} or 
\ref{fig:l5000}). 
By performing the likelihood analysis on the recently reported CBI 
\cite{astro-ph/0205384} and BIMA \cite{2002astro.ph..6012D} detections 
at the level of $15-20~\mu{\rm K}$ at $l \sim 2000-6000$ 
(figure~\ref{fig:cbibima}), we find 
$\sigma_8(\Omega_{\rm b}h/0.035)^{2/7}=1.04\pm 0.12$ at the 95\% 
statistical confidence level. 
To this we should add about 10\% systematic theoretical uncertainty due to 
the simplified modeling in our model, the numerical issues in the 
simulations, and missing physics in both. 
We should note that most of the effects we have ignored within our model 
tend to further increase $\sigma_8$: assumed fiducial value 
for $\Omega_{\rm b}h=0.035$ is somewhat high, we ignored fraction of gas 
transformed into stars, and our predictions are somewhat higher than 
the simulation results at the angular scales relevant for CBI and BIMA. 
These effects would further increase $\sigma_8$, although 
not by more than $5-10\%$. 
Such a high value for $\sigma_8$ is in tension with the primary CMB 
amplitude determination \cite{2002MNRAS.333..961L} 
and Ly-$\alpha$ forest determination 
\cite{1999ApJ...520....1C,2000ApJ...543....1M}, although it can 
be accommodated by the cluster abundance and the weak lensing results 
if $\Omega_{\rm m}\sim 0.2 \pm 0.1$. 
While the CBI experiment is surveying $1~{\rm deg^2}$ of the sky and 
still has a large instrumental noise, a somewhat larger sky coverage with 
lower noise (e.g., SZA) would significantly reduce the statistical errors 
on $\sigma_8$.
Surveys with $100~{\rm deg^2}$ such as ACT or AMIBA should be able to 
measure both the power spectrum and its covariance matrix 
(e.g. by using bootstrap sampling) to exquisite precision.

In conclusion, measurement of the angular power spectrum of 
the SZ effect offers a promising way to determine $\sigma_8$, since 
it is very sensitive to its value and is almost independent of 
$\Omega_{\rm m}$.  
It is free of observational selection effects such as
flux, surface brightness, or volume limit of the survey.
It does not require the mass of sampled haloes to be measured,  
and there is no uncertainty associated with the mass--flux calibration. 
Finally, it is within reach of the next generation of small-scale CMB 
experiments and may have already been detected by the CBI and BIMA experiments,
in which case their results imply that 
the amplitude of fluctuations at $8~h^{-1}$~Mpc scale is of order unity.

\section*{ACKNOWLEDGMENTS}

We would like to thank Antonio C. da Silva, Alexandre Refregier, 
Volker Springel, Pengjie Zhang, and their collaborators for providing 
us with their hydrodynamic simulation data results.
We would like to thank John E. Carlstrom, Carlo Contaldi, Gilbert P. Holder, 
Wayne Hu, Andrey V. Kravtsov, Ue-Li Pen, Masahiro Takada, 
and Pengjie Zhang for discussions.
U. S. acknowledges the support of NASA and Packard and Sloan 
Foundation Fellowships. 


\setcounter{section}{5}
\onecolumn


\begin{table}
 \caption[]
 {Simulation parameters. 
 The first column assigns a label to each simulation.
 The second column shows the name of a code which each simulation uses: 
 MMH ($=$Moving-Mesh Hydrodynamic code) 
 \cite{1998ApJS..115...19P},
 and RAMSES (adaptive mesh-refinement hydrodynamic code) 
 \cite{2002A&A...385..337T} are mesh codes, while
 GADGET ($=$GAlaxies with Dark matter and Gas intEracT code)
 \cite{2001NewA....6...79S}, and
 HYDRA \cite{1997NewA....2..411P} are
 SPH ($=$Smoothed-Particle Hydrodynamic) codes.
 The third column shows the number of grids for mesh codes 
 (MMH and RAMSES), or the number of SPH particles for SPH codes 
 (GADGET and HYDRA).
 The box size of simulations is 100~$h^{-1}~{\rm Mpc}$ except for
 SWH01, which uses 134~$h^{-1}~{\rm Mpc}$.
 The fourth column shows the number of two-dimensional maps which 
 each simulation has created.
 The fifth column shows the field-of-view of the simulated
 two-dimensional maps.
 Note that RKSP00 and RT00 have not created two-dimensional maps, but 
 computed the SZ angular power spectrum from the three-dimensional 
 SZ power spectrum. 
 The sixth$-$eighth columns show the cosmological parameters which
 each simulation uses. 
 All the simulations use $\Omega_{\Lambda}=1-\Omega_{\rm m}$, and 
 $n=1.0$.
 The rightmost column shows references for the simulations.
}
 \begin{center} 
  \begin{tabular}{llcccccccl}
   \hline
   Label & Code & $N_{\rm grid}$ or $N_{\rm sph}$ 
   & $N_{\rm map}$ & F.O.V.
   & $\Omega_{\rm m}$  & $\Omega_{\rm b}$ & $h$ & $\sigma_8$ 
   & Reference \\
   \hline 
   RKSP00    & MMH    & $128^3$ & --- & ---
   & 0.37 & 0.049  & 0.7  & 0.8  
   & \citeN{2000PhRvD..61l3001R}\\
   SBP01     & MMH    & $256^3$ & 12  & $2^\circ\times 2^\circ$
   & 0.37 & 0.049  & 0.7  & 0.8  
   & \citeN{2001PhRvD..630f619S}\\
   SWH01     & GADGET & $224^3$ & 15  & $1^\circ\times 1^\circ$
   & 0.3  & 0.04   & 0.67 & 0.9  
   & \citeN{2001ApJ...549..681S}\\
   daSilva01 & HYDRA  & $160^3$ & 30  & $1^\circ\times 1^\circ$
   & 0.35 & 0.0377 & 0.71 & 0.9  
   & \citeN{2001ApJ...561L..15D}\\
   RT00      & RAMSES & $256^3$ & --- & ---
   & 0.3  & 0.039  & 0.7  & 0.93 
   & \citeN{astro-ph/0012086}\\
   ZPW02     & MMH    & $512^3$ & 40  & $1^\circ\hspace{-1.1mm}.19\times 1^\circ\hspace{-1.1mm}.19$
   & 0.37 & 0.05   & 0.7  & 1.0  
   & \citeN{2002astro.ph..1375Z}\\
   \hline
  \end{tabular}
 \end{center}
 \label{tab:simulations}
\end{table}
\twocolumn


\appendix

\section[]{Gas polytropic index}

In this appendix we derive an exact formula for the 
gas polytropic index $\gamma$ and give
fitting formulae for $\gamma$ and for the mass--central
temperature normalization factor, $\eta(0)$ 
(equation~\ref{eq:Tgas0}).

We specify $\gamma$ and $\eta(0)$ uniquely by requiring that 
the gas-density profile,
$\rho_{\rm gas}(x)$, matches the dark-matter density profile, 
$\rho_{\rm dm}(x)$, in outer parts of the haloes.
We do this by solving the following equation,
\begin{equation}
 \label{eq:matching}
  s_*
  \equiv \left.\frac{d\ln \rho_{\rm gas}(x)}{d\ln x}\right|_{x=x_*}
  = \left.\frac{d\ln \rho_{\rm dm}(x)}{d\ln x}\right|_{x=x_*},
\end{equation}
where $s_*$ denotes a slope of the dark-matter density profile at 
a matching radius $x_*$.

In previous paper \cite{2001MNRAS.327.1353K} we have shown that 
equation~(\ref{eq:matching}) determines $\eta(0)$ uniquely for a
given $\gamma$,
\begin{eqnarray}
 \nonumber
  \eta(0)
  &=&
  \gamma^{-1}
  \left\{
  \left(\frac{-3}{s_*}\right)
  \left[ \frac{x_*^{-1}m(x_*)}{c^{-1}m(c)} \right]\right.\\
 \label{eq:eta0}
  & &\qquad \left. +3(\gamma-1)\left[\frac{c}{m(c)}\right]
  \int_0^{x_*} du \frac{m(u)}{u^2}
 \right\},
\end{eqnarray}
where $m(x)$ is a non-dimensional dark-matter mass profile defined by
\begin{equation}
 \label{eq:m}
  m(x) \equiv \rho_{\rm s}^{-1}\int_0^x du~u^2 \rho_{\rm dm}(u),
\end{equation}
and $\rho_{\rm s}$ is a scale density.
Equation~(\ref{eq:eta0}) makes the matching condition, 
equation~(\ref{eq:matching}), satisfied at radius $x_*$.

Next, we require that $\eta(0)$ does not depend upon a particular
choice of $x_*$; this requirement specifies $\gamma$ uniquely.
While we have done this by finding an empirical fitting formula for $\gamma$
in the previous paper (equation~25 of 
\citeNP{2001MNRAS.327.1353K}), we derive here
an exact formula for $\gamma$ by solving the following equation,
\begin{equation}
 \label{eq:etaderiv}
  \left.\frac{\partial\eta(0)}{\partial x_*}\right|_{x_*=c}=0,
\end{equation}
where $c$ is the concentration parameter (see equation~\ref{eq:c} for 
definition).
This equation means that $\eta(0)$ normalization 
does not depend upon the particular choice of $x_*$.

We find that equation~(\ref{eq:etaderiv}) gives $\gamma$ as a function
of $c$,
\begin{equation}
 \label{eq:gamma}
  \gamma
  = 1 - \left.\frac1{s_*}\right|_{x_*=c}
  +  \left.
   \frac{\partial\ln\left[m(x_*)/s_*\right]}{s_*\partial\ln x_*}
 \right|_{x_*=c}.
\end{equation}
This formula gives the polytropic index of gas in haloes
with any given dark-matter density profiles.

Hereafter, we evaluate $\gamma$ and $\eta(0)$ for the NFW profile 
\cite{1997ApJ...490..493N}, $\rho_{\rm dm}(x)=\rho_{\rm s}x^{-1}(1+x)^{-2}$.
We find 
\begin{equation}
 s_*= -\frac{1+3x_*}{1+x_*},
\end{equation}
and
\begin{equation}
 m(x)= \ln(1+x)-\frac{x}{1+x}.
\end{equation}
By substituting $s_*$ and $m(x)$ into equation~(\ref{eq:gamma}),
we obtain $\gamma$. 
We then substitute $s_*$, $m(x)$, and the derived $\gamma$ into
(\ref{eq:eta0}) to obtain $\eta(0)$.

\begin{figure}
  \begin{center}
    \leavevmode\epsfxsize=8.4cm \epsfbox{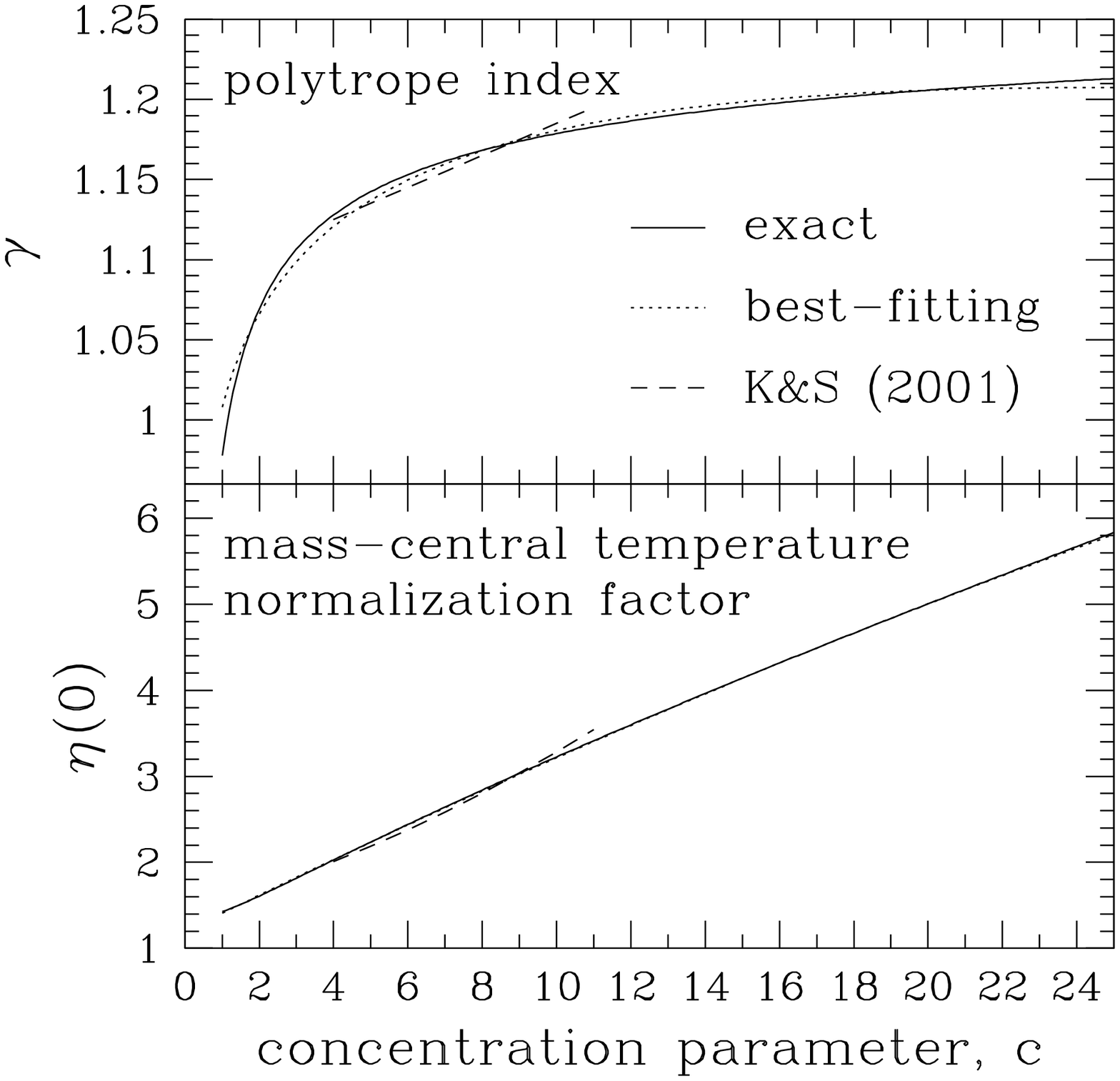}
  \end{center}
\caption[]
 {The top panel shows the polytropic index of gas in haloes, 
 $\gamma$, while the bottom panel shows the mass--central temperature 
 normalization factor, $\eta(0)$ (equation~\ref{eq:Tgas0}), 
 as a function of the concentration parameter $c$.
 The solid lines plot exact values calculated from
 equation~(\ref{eq:gamma}) and (\ref{eq:eta0}) for $\gamma$ and
 $\eta(0)$, respectively.
 The dotted lines plot the best-fitting formulae,
 equation~(\ref{eq:gammafit}) and (\ref{eq:eta0fit}) for $\gamma$ and
 $\eta(0)$, respectively.
 The dashed lines plot the fitting formulae given in
 \citeN{2001MNRAS.327.1353K} for $4<c<11$.}
\label{fig:gamma}
\end{figure}

The upper panel of figure~\ref{fig:gamma} shows $\gamma$ as a function
of $c$ compared to the fitting formula in equation (\ref{eq:gammafit}).
For comparison, the figure also plots the fitting formula which
we have given for $4<c<11$ in \citeN{2001MNRAS.327.1353K}.
The new formula, equation~(\ref{eq:gammafit}), is valid for
$1<c<25$.

By substituting $\gamma$ from  equation~(\ref{eq:gamma}) into 
equation~(\ref{eq:eta0}) we obtain $\eta(0)$ as a function of $c$.
The bottom panel of figure~\ref{fig:gamma} shows $\eta(0)$
as a function of $c$ and also the fitting formula 
$\eta(0)$ in equation (\ref{eq:eta0fit}). 
For comparison, the figure also plots the fitting formula given in 
\citeN{2001MNRAS.327.1353K}.

\label{lastpage}

\end{document}